\documentclass[review]{elsarticle}

\usepackage{multicol}
\usepackage[utf8]{inputenc}

\usepackage{textcomp}
\usepackage{siunitx}
\usepackage{comment}
\usepackage{caption}
\usepackage{subcaption}
\usepackage{blindtext}
\usepackage[toc,page,title]{appendix}
\usepackage{amsmath,amsfonts,amsthm,amssymb}
\usepackage{empheq}
\usepackage{stmaryrd}
\usepackage{mathalfa, eufrak, yfonts, mathrsfs}

\newtheorem{lemma}{Lemma}

\newtheorem{theorem}{Theorem}

\newtheorem{corollary}{Corollary}

\theoremstyle{definition}
\newtheorem{definition}{Definition}

\theoremstyle{definition}
\newtheorem*{definition*}{Definition}

\newtheorem{proposition}{Proposition}

\newtheorem{property}{Property}

\newtheorem{example}{Example}

\renewenvironment{proof}{{\bfseries Proof.}}{\qed}

\usepackage[ruled,vlined]{algorithm2e}

\usepackage[makeroom]{cancel}
\usepackage{soul}
\usepackage{pgfplots}
\usepackage{tikz}
\usetikzlibrary{decorations.pathreplacing}
\usetikzlibrary{positioning}
\usetikzlibrary{arrows}
\usetikzlibrary{arrows.meta}
\usetikzlibrary{shapes.geometric}
\usetikzlibrary{shapes.misc}
\tikzset{cross/.style={cross out, draw=red, minimum size=40*(#1-\pgflinewidth), inner sep=0pt, outer sep=0pt},
	cross/.default={1pt}}

\makeatletter
\newcommand{\osymbol}[1]{\mathbin{\mathpalette\make@circled#1}}
\newcommand{\make@circled}[2]{
	\ooalign{$\m@th#1\smallbigcirc{#1}$\cr\hidewidth$\m@th#1#2$\hidewidth\cr}
}
\newcommand{\smallbigcirc}[1]{
	\vcenter{\hbox{\scalebox{1}{$\m@th#1\bigcirc$}}}
}
\makeatother

\usepackage{graphicx}
\usepackage{lineno,hyperref}
\modulolinenumbers[5]

\journal{Information Sciences}

\begin{document}

\begin{frontmatter}

\title{
	Focal points and their implications for M\"obius Transforms and 	Dempster-Shafer Theory\tnoteref{mytitlenote}}
\tnotetext[mytitlenote]{This work was carried out and co-funded in the framework of the Labex MS2T and the Hauts-de-France region of France. It was supported by the French Government, through the program ``Investments for the future'' managed by the National Agency for Research (Reference ANR-11-IDEX-0004-02).}
\author{Maxime Chaveroche$^*$, Franck Davoine, V\'eronique Cherfaoui}
\ead{name.surname@hds.utc.fr}
\address{Sorbonne University Alliance, Universit\'e de technologie de Compi\`egne, CNRS, Heudiasyc,\\
	CS 60319 - 60203 Compi\`egne Cedex, France}
\cortext[mycorrespondingauthor]{Corresponding author}
\begin{abstract}
Dempster-Shafer Theory (DST) generalizes Bayesian probability theory, offering useful additional information, but suffers from a much higher computational burden. A lot of work has been done to reduce the time complexity of information fusion with Dempster's rule, which is a pointwise multiplication of two zeta transforms, and optimal general algorithms have been found to get the complete definition of these transforms. Yet, it is shown in this paper that the zeta transform and its inverse, the M\"obius transform, can be exactly simplified, fitting the quantity of information contained in belief functions. Beyond that, this simplification actually works for any function on any partially ordered set. It relies on a new notion that we call \textit{focal point} and that constitutes the smallest domain on which both the zeta and M\"obius transforms can be defined. We demonstrate the interest of these general results for DST, not only for the reduction in complexity of most transformations between belief representations and their fusion, but also for theoretical purposes. Indeed, we provide a new generalization of the conjunctive decomposition of evidence and formulas uncovering how each decomposition weight is tied to the corresponding mass function.
\end{abstract}

\end{frontmatter}


\section{\textbf{Introduction}}\label{sec:introduction}

Dempster-Shafer Theory (DST) \cite{shafer76} is an elegant formalism that generalizes Bayesian probability theory by considering the specificity of evidence. This means that it 
enables a source (e.g. some sensor model) to represent its belief in the state of a variable by assigning credit not only directly to a possible state, as in Bayesian probability theory, but also to groups of states, reflecting some model uncertainty or randomness. This assignment of belief is called a \textit{mass function}.
It generalizes the notion of probability distribution by providing meta-information that quantifies the level of uncertainty about one's believes considering the way one established them, which is critical for decision making.

Nevertheless, this information comes with a cost: let $\Omega$ be the set containing all possible states. This set $\Omega$ has $2^{|\Omega|}$ subsets. Thus, in DST, we consider $2^{|\Omega|}$ potential values instead of only $|\Omega|$ in Bayesian probability theory, which can lead to computationally and spatially very expensive algorithms. Computations can become difficult to perform for more than a dozen possible states (e.g. 20 states in $\Omega$ generate more than a million subsets), although we may need to consider much more of them
(e.g. for classification or identification). 
This imposes a choice in DST between expressiveness of the model (i.e. $|\Omega|$) and fast computations (especially considering limited resources such as in embedded systems). 
To tackle this issue, a lot of work has been done to reduce the complexity of transformations used to combine belief sources with Dempster's rule \cite{dempster68}. We distinguish between two approaches that we call \textit{powerset-based} and \textit{evidence-based}.

The \textit{powerset-based} approach concerns all algorithms based on the structure of the powerset $2^{\Omega}$. 
They have a complexity dependent on $|\Omega|$. Early works \cite{barnett81,gordon85,shenoy86,shafer87} proposed optimizations by restricting the structure of evidence to only singletons and their negation, which greatly restrains the expressiveness of DST. Later, a family of optimal algorithms working in the general case, that is, those based on the \textit{Fast M\"obius Transform} (FMT) \cite{FMT}, was discovered. Their complexity is $O(|\Omega|. 2^{|\Omega|})$ in time and $O(2^{|\Omega|})$ in space. The FMT has become the de facto standard for the computation of every transformation in DST. Consequently, efforts were made to reduce the size of $\Omega$ to benefit from the optimal algorithms of the FMT. More specifically, \cite{wilson2000} refers to the process of conditioning by the \textit{combined core} (intersection of the unions of all sets of nonnull mass
of each belief source) and \textit{lossless coarsening} (partitioning of $\Omega$ into super-elements that each represents a group of elements always appearing together in all sets of nonnull mass). There are also approximation methods such as Monte Carlo methods \cite{wilson2000}, which depend on a number of trials that must be large and grows with $|\Omega |$, and lossy coarsening \cite{denoeux2002approximating}, but we focus here on exact methods. More recently, DST has been generalized to any lattice (not just $2^\Omega$) in  \cite{grabisch2009belief}, which has been used to propose a method \cite{denoeux2010dempster} limiting its expressiveness to only intervals in $\Omega$ in order to keep DST transformations tractable. If there is no order between the elements of $\Omega$, this method comes down again to restricting the assignment of credit to only singletons.

The \textit{evidence-based} approach concerns all algorithms that aim to restrict computations (not expressivity) to the only subsets that have a nonnull mass, i.e. that contain information (\textit{evidence}). These are called \textit{focal sets} and are usually far less numerous than $2^{|\Omega|}$. This approach, also refered as the \textit{obvious} one, implicitly originates from the seminal work of Shafer \cite{shafer76} and is often more efficient than the powerset-based one since it only depends on information contained in sources in a quadratic way. Doing so, it allows for the exploitation of the full potential of DST by enabling one to choose any frame of discernment, without concern about its size. Moreover, the evidence-based approach benefits directly from the use of approximation methods, some of which are very efficient \cite{sarabi-jamab16}. Therefore, this approach seems superior to the FMT in most use cases, above all when $|\Omega|$ is large, where an algorithm with exponential complexity is just intractable.

But, unfortunately, focal sets are not sufficient to define the zeta and M\"obius transforms. 
In particular, if one wishes to compute the multiplicative M\"obius transform of an additive zeta transform (e.g. computing the conjunctive or disjunctive weight function from the commonality or implicability function), 
focal sets are not enough in the general case. 
For this, we already proposed in \cite{me_gretsi}\footnote{This short reference is written in french, but does not need to be read to follow this article.} the notion of \textit{focal point}, which is sufficient to completely define the conjunctive and disjunctive decompositions.

One other argument against the evidence-based approach is the lack of knowledge about the structure of the set of focal sets due to its inconstancy. Indeed, the FMT draws its power from the knowledge of the structure of Boolean lattices. Doing so, all algorithm using only focal sets is forced to have a quadratic complexity in the number of these focal sets, which can be worse than the complexity of FMT algorithms when this number approaches the size of the powerset $2^\Omega$. Yet, \textit{focal points} do have a specific structure that has been successfully exploited in \cite{me}, where we proposed methods with complexities that are variable but always inferior to the complexity of the FMT, its trimmed version \cite{bjorklundtrimmed} and more generally the Fast Zeta Transform \cite{bjorklund2016fast, kaski2016fast}, and may be even lower than $O(|\Omega|.F)$ in some cases, where $F$ is the number of focal sets.

This present article extends our previous works \cite{me_gretsi} and \cite{me}, focusing on the study of these focal points to provide their generalization to any partially ordered set and the reformulation of the M\"obius inversion theorem based on them, which were missing. This new contribution also applies to the multiplication of any function by the zeta or M\"obius function in any incidence algebra \cite{rota}. The second part of this article proposes applications in DST exploiting these focal points. We limit this second part to the classical DST, for the sake of clarity, although our results are applicable to its generalization \cite{grabisch2009belief} to any lattice. 

This paper is organized as follows: Section \ref{preliminaries} presents the elements on which our notions are built. Section \ref{methods} presents our contributions to the zeta and M\"obius transforms. Section \ref{applications} discusses multiple applications to DST. Finally, we conclude this article in section \ref{conclusion}.

\section{\textbf{Background of our method}}\label{preliminaries}

Let $(P, \leq)$ be a semifinite (lower semifinite for $\leq$, upper semifinite for $\geq$) set partially ordered by some binary operator noted $\leq$, e.g. the powerset $(2^\Omega, \subseteq)$ or $(2^\Omega, \supseteq)$ containing all subsets of a set $\Omega$ ordered by inclusion, the set $(\mathbb{N}^*, |)$ of all positive integers ordered by divisibility, the set $(\mathbb{N}, \leq)$ of all nonnegative integers ordered by $\leq$, etc.

\subsection{\textbf{Zeta transform (``Discrete integral'')}} The zeta transform $g: P \rightarrow \mathbb{R}$ of a function $f: P \rightarrow \mathbb{R}$ is defined as follows:
\begin{align}\label{zeta_g}
\forall y \in P,\quad g(y) = \sum_{x \leq y} f(x)
\end{align}
It is \textbf{analogous to integration} in a discrete domain.
Its name comes from incidence algebra \cite{rota}, in which it corresponds to the multiplication of $f:P^2 \rightarrow \mathbb{R}$ with the \textit{zeta function} $\zeta:P^2 \rightarrow \{0,1 \}$, such that $\zeta(x,y)=1$ if $x\leq y$, i.e. 
\begin{align*}
\forall a,b \in P,\quad (f * \zeta)(a,b) =  \sum_{a \leq x \leq b} f(a,x).\zeta(x,b) = \sum_{a \leq x \leq b} f(a,x) = g(a,b)
\end{align*}

\begin{example}\label{m_to_b}
	In DST, the implicability function $b$ is defined as the zeta transform of the mass function $m$ in $(2^\Omega, \subseteq)$, i.e. 
	$\displaystyle\forall y \in 2^\Omega,\quad b(y) = \sum_{x \subseteq y} m(x)$.
\end{example}

\begin{example}\label{m_to_q}
	In DST, the commonality function $q$ is defined as the zeta transform of the mass function $m$ in $(2^\Omega, \supseteq)$, i.e. $\displaystyle\forall y \in 2^\Omega,\quad q(y) = \sum_{x \supseteq y} m(x)$.
\end{example}

\subsection{\textbf{M\"obius transform (``Discrete derivative'')}} The M\"obius transform of $g$ is $f$. It is \textbf{analogous to differentiation} in a discrete domain and is defined by the \textit{M\"obius inversion formula}:
\begin{align}\label{mob_trans}
\forall y \in P,\quad f(y) = \sum_{x \leq y} g(x) . \mu^{\phantom{\dagger}}_{P, \leq}(x,y)
\end{align}
where $\mu^{\phantom{\dagger}}_{P, \leq}$ is the M\"obius function of $(P, \leq)$, defined in its general form 
in \cite{rota} as follows:
\begin{align}\label{mob_func_sum}
\forall x,y \in P,\quad \sum_{x \leq z \leq y} \mu^{\phantom{\dagger}}_{P, \leq}(x, z) = \sum_{x \leq z \leq y} \mu^{\phantom{\dagger}}_{P, \leq}(z, y) = 0,
\end{align}
with $\mu^{\phantom{\dagger}}_{P,\leq}(x,x) = 1$. This can be rewritten in the following recursive form:
\begin{align}\label{mobius_func}
\forall x,y \in P,\quad \mu^{\phantom{\dagger}}_{P, \leq}(x,y) = 
\begin{cases}
1	&\text{ if $x = y$}\\
- \displaystyle\sum_{x < z \leq y} \mu^{\phantom{\dagger}}_{P, \leq}(z, y)	&\text{ otherwise}
\end{cases}
\end{align}
In an incidence algebra, the M\"obius transform of $g$ corresponds to the multiplication of $g:P^2 \rightarrow \mathbb{R}$ with the M\"obius function $\mu$, i.e. 
\begin{align*}
\forall a,b \in P,\quad (g * \mu^{\phantom{\dagger}}_{P, \leq})(a,b) =  \sum_{a \leq x \leq b} g(a,x).\mu^{\phantom{\dagger}}_{P, \leq}(x,b) = f(a,b)
\end{align*}

\begin{example}\label{b_to_m}
	Taking back the functions $m$ and $b$ from \autoref{m_to_b}, $m$ is the M\"obius transform of $b$ in $(2^\Omega, \subseteq)$, i.e. $\forall y \in 2^\Omega,\quad m(y) = \displaystyle\sum_{x \subseteq y} b(x)~.~ (-1)^{|y| - |x|}$.
\end{example}

\begin{example}\label{q_to_m}
	Taking back the functions $m$ and $q$ from \autoref{m_to_q}, $m$ is the M\"obius transform of $q$ in $(2^\Omega, \supseteq)$, i.e. $\forall y \in 2^\Omega,\quad m(y) = \displaystyle\sum_{x \supseteq y} q(x)~.~ (-1)^{|y| - |x|}$.
\end{example}

\subsection{\textbf{Multiplicative M\"obius inversion theorem}}

There is also a multiplicative version of the zeta and M\"obius transforms with the same properties in which the sum is replaced by a product:
\begin{align*}
\forall y \in P,\quad g(y) &= \prod_{x \leq y} f(x)\nonumber
\quad\Leftrightarrow\quad f(y) = \prod_{\substack{x \leq y}}~
g(x)^{\mu^{\phantom{\dagger}}_{P,\leq}(x,y)}
\end{align*}

\begin{example}\label{b_to_v}
In DST, the disjunctive weight function $v$ is defined as the inverse of the multiplicative M\"obius transform of $b$ from \autoref{m_to_b} in $(2^\Omega, \subseteq)$, i.e. $\forall y \in 2^\Omega,\quad v(y) = \displaystyle\prod_{x \subseteq y} b(x)^{-(-1)^{|y| - |x|}}$.
\end{example}

\begin{example}\label{q_to_w}
	In DST, the conjunctive weight function $w$ is defined as the inverse of the multiplicative M\"obius transform of $q$ from \autoref{m_to_q} in $(2^\Omega, \supseteq)$, i.e. $\forall y \in 2^\Omega,\quad w(y) = \displaystyle\prod_{x \supseteq y} q(x)^{-(-1)^{|y| - |x|}}$.
\end{example}

\newcommand{\supp}[1]{\sloppy\text{supp$(#1)$}}
\subsection{\textbf{Support of a function in $P$}} The support $\supp{f}$ of a function $f: P \rightarrow \mathbb{R}$ is defined as $\text{supp}(f) = \{ x \in P ~/~ f(x) \neq 0 \}$. 
For example, in DST, the set of focal elements of a mass function $m$ is $\text{supp}(m)$.

{With this notion, it is obvious that Eq. \ref{zeta_g} can be reduced to:
\begin{align}\label{compact_zeta}
\forall y \in P,\quad g(y) = \sum_{\substack{x\in \supp{f}\\x \leq y}} f(x)
\end{align}}
\subsection{\textbf{Order theory}}

\paragraph{\textbf{Minimal/Maximal elements}} Minimal elements of a set $S$ are its elements such that there is no element in $S$ that is less than them. The set containing them is noted $\min(S)$. If there is only one minimal element in $S$, it is its \textit{minimum}. For example, in a totally ordered set (i.e. a \textit{chain}), there can only be one minimal element, i.e. its minimum. Dually, the same principle holds for maximal elements of a set $S$, noted $\max(S)$, and its \textit{maximum}.

\paragraph{\textbf{Supremum/Infimum}} The \textit{supremum} (also known as \textit{join}) of a set $S$ of elements of $P$ is its least upper bound, i.e. the least element in $P$ that is greater than or equal to every element of $S$. It is noted $\bigvee S$, but only exists if the set of the upper bounds of $S$ has only one minimal element. In particular, if $S$ has a maximum, then it is $\bigvee S$. Dually, the \textit{infimum} (also known as \textit{meet}) of a set $S$ of elements of $P$ is its greatest lower bound, i.e. the greatest element in $P$ that is less than or equal to every element of $S$. It is noted $\bigwedge S$.

\paragraph{\textbf{Lattice and semilattice}} An \textit{upper semilattice} $S$ is a set such that every of its nonempty subsets has its supremum in $S$. A \textit{lower semilattice} $S$ is a set such that every of its nonempty subsets has its infimum in $S$. A \textit{lattice} is both an upper and a lower semilattice.
In particular, all element of a \textit{finite} lattice $L$ can be described as either $\bigwedge L$ or the supremum (join) of a nonempty subset of its \textit{join-irreducible elements}.

\newcommand{\supfp}[1]{\sloppy\text{${^\vee\supp{#1}}$}}
\newcommand{\supfpof}[1]{\sloppy\text{${^\vee#1}$}}
\newcommand{\inffp}[1]{\sloppy\text{${^\wedge\supp{#1}}$}}
\newcommand{\inffpof}[1]{\sloppy\text{${^\wedge#1}$}}

\section{\textbf{Focal points and our \textit{Efficient} M\"obius inversion formula}}\label{methods}

The purpose of this section is to present our work on what we call the \textit{Efficient M\"obius inversion formula} and its developments. 
Section \ref{intuition} gives an overview of the approach. Section \ref{etmf} tries to simplify the M\"obius inversion formula of Eq. \ref{mob_trans}, highlighting the emergence of our focal points. Section \ref{fp_section} properly defines these focal points and exposes the simplifications they allow on the M\"obius inversion formula. Section \ref{fp_computation} proposes ways to compute them. Finally, section \ref{link_section} exploits one of these ways to uncover links between the additive and multiplicative M\"obius transforms of a same function. In addition, section \ref{discussions} discusses some aspects of our approach and section \ref{theory_practice} bridges the gap between our theoretical results and their usage.

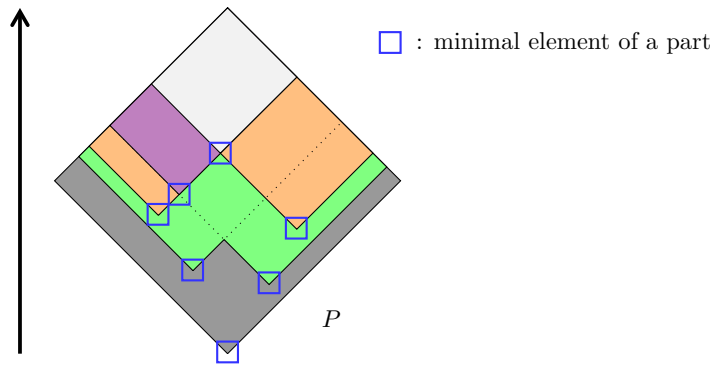
\begin{figure}[ht]
	\centering
	\hspace{2cm}
	\begin{tikzpicture}[scale=0.92, every node/.style={transform shape},
	node/.style={draw, dot,minimum size=0.2cm, inner sep=0pt},
	det/.style={draw, diamond,minimum size=1.1cm, inner sep=0pt},
	rect/.style={draw, rectangle,minimum size=1.1cm, inner sep=2pt}
	]
	
	\filldraw[fill=white!20!gray, draw=black] (0,0) -- (2.5, 2.5) -- (0,5) -- (-2.5,2.5) -- (0,0);
	\filldraw[fill=white!50!green, draw=black] (-0.5,1.2) -- (-2.15, 2.85) -- (0,5) -- (2.3, 2.7) -- (0.6, 1) -- (-0.05, 1.65) -- (-0.5,1.2);
	
	
	\filldraw[fill=white!50!orange,] (-1,2) -- (-2, 3) -- (0,5) -- (2.1, 2.9) -- (1,1.8) -- (-0.105, 2.9) -- (-1,2);
	\draw[dotted] (-1.7, 3.3) -- (-0.05, 1.65) -- (1.65, 3.35);
	\filldraw[fill=white!50!violet] (-0.7, 2.3) -- (-1.7, 3.3) -- (0, 5) -- (1, 4) -- (-0.7, 2.3);
	\filldraw[
	fill=white!90!gray, 
	draw=black] (-0.105, 2.9) -- (-1.1, 3.9) -- (0,5) -- (1,4) -- (-0.105, 2.9);
	\node (P) at (1.5, 0.5) {$P$};
	\draw[line width=0.05cm,-{angle 60[length=100mm, width=2000mm]}] (-3,0) -- (-3., 5);
	\node[draw,thick,white!20!blue,rectangle,minimum size=0.3cm,inner sep=0pt] (afp) at (-0.505, 1.21) {};
	\node[draw,thick,white!20!blue,rectangle,minimum size=0.3cm,inner sep=0pt] (afp) at (-1.005, 2.015) {};
	\node[draw,thick,white!20!blue,rectangle,minimum size=0.3cm,inner sep=0pt] (afp) at (0.9925, 1.815) {};
	\node[draw,thick,white!20!blue,rectangle,minimum size=0.3cm,inner sep=0pt] (afp) at (-0.105, 2.9) {};
	\node[draw,thick,white!20!blue,rectangle,minimum size=0.3cm,inner sep=0pt] (afp) at (0.6, 1.025) {};
	\node[draw,thick,white!20!blue,rectangle,minimum size=0.3cm,inner sep=0pt] (afp) at (-0.7, 2.305) {};
	\node[draw,thick,white!20!blue,rectangle,minimum size=0.3cm,inner sep=0pt] (afp) at (2.35,4.5) {};
	\node[draw,thick,white!20!blue,rectangle,minimum size=0.3cm,inner sep=0pt] (afp) at (0, 0.025) {};
	\node (l2desc) at (4.855,4.5) {: minimal element of a part};
	\end{tikzpicture}
	\caption{\small
		Example of image partition $\mathcal{G}$ of $P$ corresponding to some function $g$. The (discretized) area of the diamond represents $P$, which has a minimum in this example. Each point of it corresponds to one of its elements. The arrow on the left represents the order $\leq$. 
		Two points are ordered only if the upward vector aligning them makes an angle between \ang{-45} and \ang{45} with the big arrow vector on the left. If so, 
		then, of the two points, the one that is lower in the figure is lower in $P$. 
		Each color represents an image through $g$. 
		Each part $X$ of this partition $\mathcal{G}$ contains elements of same image through $g$ and can be described as all elements between $\min(X)$ and $\max(X)$. In this example, each color also represents a part. We see that $g$ can be defined in a compact way as a list or tree of parts, each described by its minimal and maximal elements. 
	}\label{fig:image_partition}
\end{figure}

Let us note $(P, \leq)$ the 
set $P$ partially ordered by some binary operator noted $\leq$, and let $g$ and $f$ be the functions of Eq. \ref{zeta_g}. 
{For the sake of simplicity, we will consider in this section that $P$ is finite. If $P$ is semifinite or infinite, see section \ref{discussions}.  We assume that $f$ is defined in a compact way, simply through $\supp{f}$. We also assume that $g$ is defined in a compact way, through a partition of $P$, noted $\mathcal{G}$, into parts $X$ such that all elements of $P$ greater than at least one element of $\min(X)$ and less than at least one element of $\max(X)$ is in $X$ and has the same image through $g$ as any other element of $X$.
Multiple partitions may fullfill these conditions. For instance, we could force these parts to be intervals (i.e. to force $|\min(X)|=|\max(X)|=1$ for all part $X$) or not. This is of no importance for what follows, but the fewer parts there are, the fewer computations will be needed. An example of such an \textit{image partition} is illustrated in Fig. \ref{fig:image_partition}. The image through $g$ of all elements in $P$ is thus determined by only minimal and maximal elements of parts.
Also, note that all of our following results can be applied to any incidence algebra, simply considering $g(a, \cdot)$ and the support $\text{supp}(f(a,\cdot)) = \{ x \in P ~/~ f(a,x) \neq 0 \}$ for some $a\in P$ instead of $\text{supp}(f)$, since functions are then defined on intervals instead of single elements. 
}

\subsection{\textbf{Problem statement and intuition}}\label{intuition}

Let us start by translating the problems stated in introduction of this article into formal terms.
As showed in Eq. \ref{compact_zeta}, $g$ can be computed from $\supp{f}$ only. However, $g$ cannot be completely defined in the general case from $g(\supp{f})$ alone. Furthermore, it is not possible to determine $\supp{f}$
from the definition of $g$ before computing $f$. These two issues prevent us from limiting computations so that it scales linearly with the quantity of information in these transforms, i.e. $|\supp{f}|$. Nevertheless, it is possible to determine the smallest set containing both $\supp{f}$ and all the defining elements of $P$ for $g$, either from $f$ or from $g$. 

Giving some reduced common superset for these has been done in the past: For conjunctive fusion in DST, one can find in \cite{wilson2000} that computations can be limited to the powerset of $\supp{f_1} \cap \supp{f_2}$, where $f_1$ and $f_2$ are two mass functions to combine. In \cite{bjorklundtrimmed}, a method was proposed to limit computations of the zeta transform to the elements of the powerset that are greater than an element of $\min(\supp{f})$. In contrast, here we propose the smallest common superset for $f$ and $g$ in any semilattice. To the best of our knowledge, our approach is also the first one to propose a reduced common superset based on $g$ alone for the computation of its M\"obius transform (i.e. $f$).
The elements of this smallest common superset are what we call \textit{the focal points of $f$ and $g$}. 

The idea is simple: track the influence of $f$ on $g$ by looking at elements from $P$ that are greater than the same elements from $\supp{f}$, since they are necessarily associated with the same image through $g$, and select one \textit{representative} for each of these configurations.  Computing the zeta transform of $f$ on these \textit{representatives} only will yield the complete definition of $g$. Then, use the M\"obius function associated with the partially ordered set made of these \textit{representatives} to compute the M\"obius transform of $g$ and get the complete definition of $f$.

These \textit{representatives} are our focal points. By definition, their image through $g$ contains all possible images through $g$, excepted 0 under some conditions. The same can be stated with $f$.
However, several questions arise: How to get the images of all other elements of $P$? Is there a best \textit{representative} and how to select it for each of these subsets of $\supp{f}$? See section \ref{fp_section}. How to find them efficiently and how to find them without knowing $\supp{f}$? See section \ref{fp_computation}. Is the M\"obius transform for these \textit{representatives} really the same in this reduced set as in $P$? See section \ref{etmf}. 

\newcommand{\suppart}{\sloppy\text{$\mathcal{P}_{/(\supp{f}, \leq)}$}}
\newcommand{\suppartof}[1]{\sloppy\text{$\mathcal{P}_{/(#1, \leq)}$}}
\newcommand{\suppartdual}{\sloppy\text{$\mathcal{P}_{/(\supp{f}, \geq)}$}}
\newcommand{\suppartdualof}[1]{\sloppy\text{$\mathcal{P}_{/(#1, \geq)}$}}
\newcommand{\lowerclosure}[1]{\sloppy\text{${_{\supp{f}}^{\downarrow} #1}$}}
\newcommand{\lowerclosurein}[2]{\sloppy\text{${_{#1}^{\downarrow} #2}$}}
\newcommand{\upperclosure}[1]{\sloppy\text{${^{\supp{f}}_{\uparrow} #1}$}}
\newcommand{\upperclosurein}[2]{\sloppy\text{${^{#1}_{\uparrow} #2}$}}

\begin{figure}[t]
	\centering
	\begin{subfigure}{0.5\textwidth}
		\centering
		\begin{tikzpicture}[scale=0.79, every node/.style={transform shape},
		node/.style={draw, dot,minimum size=0.2cm, inner sep=0pt},
		det/.style={draw, diamond,minimum size=1.1cm, inner sep=0pt},
		rect/.style={draw, rectangle,minimum size=1.1cm, inner sep=2pt}
		]
		
		\filldraw[fill=white!20!gray, draw=black] (0,0) -- (2.5, 2.5) -- (0,5) -- (-2.5,2.5) -- (0,0);
		\filldraw[fill=white!50!green, draw=black] (-0.5,1.2) -- (-2.15, 2.85) -- (0,5) -- (1.65, 3.35) -- (-0.5,1.2);
		\node (a) at (-0.5, 1.2) {$\bullet$};
		
		\filldraw[fill=white!50!orange, draw=black] (-1,2) -- (-2, 3) -- (0,5) -- (1,4) -- (-1,2);
		\node (b) at (-1, 2) {$\bullet$};
		\filldraw[fill=white!50!blue, draw=black] (1,1.8) -- (-1.1, 3.9) -- (0,5) -- (2.1, 2.9) -- (1,1.8);
		\node (c) at (1, 1.8) {$\bullet$};
		\filldraw[
		fill=white!50!violet, 
		draw=black] (0.55, 2.25) -- (-1.1, 3.9) -- (0,5) -- (1.65, 3.35) -- (0.55, 2.25);
		\filldraw[
		fill=white!90!gray, 
		draw=black] (-0.105, 2.9) -- (-1.1, 3.9) -- (0,5) -- (1,4) -- (-0.105, 2.9);
		\node (P) at (1.5, 0.5) {$P$};	
		\node (s1) at (-0.5, 0.95) {$s_2$};
		\node (s2) at (-0.8, 1.8) {$s_1$};
		\node (s3) at (0.9, 1.6) {$s_3$};
		\node (ymark) at (0.25, 4) {$\times$};	
		\node (y) at (0.35, 4.25) {$y$};
		\draw[dashed] (0.25,4) -- (-1.9, 1.85) -- (0,-0.05) -- (2.15, 2.1) -- (0.25,4);
		\draw[line width=0.05cm,-{angle 60[length=100mm, width=2000mm]}] (-3,0) -- (-3., 5);
		\node (l1) at (2,5.5) {$\bullet$ : element of $S$};
		\end{tikzpicture}
		\caption{\smaller Level partition}
		\label{fig:level_partition}
	\end{subfigure}%
	\begin{subfigure}{0.5\textwidth}
		\centering
		\begin{tikzpicture}[scale=0.79, every node/.style={transform shape},
		node/.style={draw, dot,minimum size=0.2cm, inner sep=0pt},
		det/.style={draw, diamond,minimum size=1.1cm, inner sep=0pt},
		rect/.style={draw, rectangle,minimum size=1.1cm, inner sep=2pt}
		]
		
		\filldraw[fill=white!20!gray, draw=black] (0,0) -- (2.5, 2.5) -- (0,5) -- (-2.5,2.5) -- (0,0);
		\filldraw[fill=white!50!green, draw=black] (-0.5,1.2) -- (-2.15, 2.85) -- (0,5) -- (1.65, 3.35) -- (-0.5,1.2);
		\node (a) at (-0.5, 1.2) {$\bullet$};
		
		\filldraw[fill=white!50!orange, draw=black] (-1,2) -- (-2, 3) -- (0,5) -- (1,4) -- (-1,2);
		\node (b) at (-1, 2) {$\bullet$};
		\filldraw[fill=white!50!blue, draw=black] (1,1.8) -- (-1.1, 3.9) -- (0,5) -- (2.1, 2.9) -- (1,1.8);
		\node (c) at (1, 1.8) {$\bullet$};
		\filldraw[
		fill=white!50!violet, 
		draw=black] (0.55, 2.25) -- (-1.1, 3.9) -- (0,5) -- (1.65, 3.35) -- (0.55, 2.25);
		\filldraw[
		fill=white!90!gray, 
		draw=black] (-0.105, 2.9) -- (-1.1, 3.9) -- (0,5) -- (1,4) -- (-0.105, 2.9);
		\node (P) at (1.5, 0.5) {$P$};	
		\node (s1) at (-0.5, 0.95) {$s_2$};
		\node (s2) at (-0.8, 1.8) {$s_1$};
		\node (s3) at (0.9, 1.6) {$s_3$};
		\node (ymark) at (0.25, 4) {$\times$};	
		\node (y) at (0.35, 4.25) {$y$};
		\draw[dashed] (0.25,4) -- (-1.9, 1.85) -- (0,-0.05) -- (2.15, 2.1) -- (0.25,4);
		\draw[line width=0.05cm,-{angle 60[length=100mm, width=2000mm]}] (-3,0) -- (-3., 5);
		\node[draw,thick,white!20!red,circle,minimum size=0.25cm,inner sep=0pt] (afp) at (-0.505, 1.21) {};
		\node[draw,thick,white!20!red,circle,minimum size=0.25cm,inner sep=0pt] (bfp) at (-1.005, 2.015) {};
		\node[draw,thick,white!20!red,circle,minimum size=0.25cm,inner sep=0pt] (cfp) at (0.9925, 1.815) {};
		\node[draw,thick,white!20!red,circle,minimum size=0.25cm,inner sep=0pt] (acfp) at (0.55, 2.25) {};
		\node[draw,thick,white!20!red,circle,minimum size=0.25cm,inner sep=0pt] (abfp) at (-0.105, 2.9) {};
		\node (l1) at (2,5.5) {$\bullet$ : element of $S$};
		\node[draw,thick,white!20!red,circle,minimum size=0.25cm,inner sep=0pt] (l2) at (0.84,5) {};
		\node (l2desc) at (2.255,5) {: element of $\supfpof{S}$};
		\end{tikzpicture}
		\caption{\smaller Join-closure of $S$}
		\label{fig:focal_points}
	\end{subfigure}
	\caption{\small Same format as in Fig \ref{fig:image_partition}. (a) Illustration of the concept of level partition. Each color represents a part of the partition $\suppartof{S}$, where $S= \{ s_1, s_2, s_3 \}$. In a level partition, every part $X$ is made of elements of same lower closure in $S$, i.e. $(\downarrow X) \cap S$. The part delimited by a dashed contour is the lower closure of some element $y$ in $P$, i.e. $\downarrow y$. We see that all elements of $S$ are in the lower closure of $y$. So, the lower closure in $S$ of $y$ is $S$. (b) Projection of all the elements of the join-closure $\supfpof{S}$ onto the level partition of (a), assuming each part $X$ of $\suppartof{S}$ where $X \subseteq {\uparrow S}$ (i.e. all parts except the gray one) has a minimum (no assumption on $\min(P)$ is made).}
\end{figure}
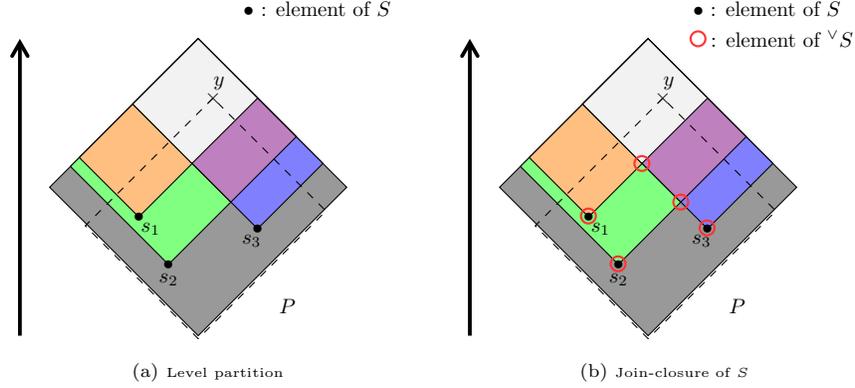

\subsection{\textbf{Simplifying the M\"obius inversion formula}}\label{etmf}

Let us start by showing that these focal points are unique and arise naturally in the expression of the M\"obius transform defined in $P$. 
In the following, we will mainly focus on the additive M\"obius inversion formula since it displays the same properties as the multiplicative form (think of the multiplicative form as the exponential of the additive form). In the end, our study exploits the neutrality of some values in the sum or product. For the addition, the neutral element is 0, hence our use of the support \supp{f}. To transpose our findings to the multiplicative form, consider instead \supp{f-1},
as the neutral element for the multiplication is 1.

\newpage
\begin{definition}[\textit{Level partition}]\label{suppart_def}
	For any subset $S \subseteq P$, let us refer to the elements of $S$ less than some element $x \in P$ as the \textit{lower closure} ${{(\downarrow x)} \cap S}$ of $x$ in $S$, i.e. ${{(\downarrow x)} \cap S} = \{ s \in S ~/~ s \leq x \}$. 
	{In accordance with convention, let us also use the notation ${{(\downarrow X)} \cap S} = \{ s \in S ~/~ \exists x \in X,~ s \leq x \}$, where $X \subseteq P$.}
	
	{For conciseness, we will use the aliases $\lowerclosurein{S}{x} = {{(\downarrow x)} \cap S}$ and $\upperclosurein{S}{x} = {{(\uparrow x)} \cap S}$ in the following, where $x$ can be a subset of $P$ or one of its elements.}
	
	We define a \textit{level partition}, noted \suppartof{S} and read as \textit{$P$ divided according to $S$ and the order $\leq$}, as the partition of $P$ such that for any distinct parts $X,Y \in \suppartof{S}$, we have $\lowerclosurein{S}{X} \neq \lowerclosurein{S}{Y}$ and such that for any part $X \in \suppartof{S}$, any element $x \in X$ satisfies $\lowerclosurein{S}{x} = \lowerclosurein{S}{X}$. 
	We say that all elements in $X$ are at the same \textit{level} regarding the elements of $S$ and the order $\leq$. We would have used the \textit{upper closure} for the dual order $\geq$. This concept is illustrated in Fig. \ref{fig:level_partition}. To summarize, we have:
	\begin{align*}
	{\suppartof{S} = \left\lbrace~ X \subseteq P ~/~ X \neq \emptyset,~ \forall y \in P,\quad~
	\left(\forall x \in X,~ \lowerclosurein{S}{x} = \lowerclosurein{S}{y}\right) \Leftrightarrow y \in X ~\right\rbrace}
	\end{align*}
	
\end{definition}

{In particular, one may notice that $\suppartof{\supp{f}}$ is also an image partition of $P$ with respect to $g$. }

\begin{definition}[\textit{M\"obius function aggregate}]\label{eta_def}
	For any element $y \in P$, for any nonempty set of elements $S \subseteq P$ and for any part $X$ of the partition $\suppartof{S}$, we define our M\"obius function \textit{aggregate} $\eta$ as follows:
	\begin{align*}
	{\eta^{\phantom{\dagger}}_{S,\leq, P}(X, y) =
	\sum_{\substack{z \in X\\z \leq y}}\mu^{\phantom{\dagger}}_{P,\leq}(z,y)}
	\end{align*}
	Notice that if $y \in \min(X)$, then $\eta^{\phantom{\dagger}}_{S,\leq, P}(X, y) = \mu^{\phantom{\dagger}}_{P,\leq}(y,y) = 1$, according to Eq. \ref{mobius_func}. The notation $\eta^{\phantom{\dagger}}_{S,\leq, P}$ is similar to $\suppartof{S}$ and is read as \textit{$\eta$ given $S$ and the elements of $P$ above it}, where the term \textit{above} depends on the order $\leq$.
\end{definition}

{Thanks to \autoref{suppart_def} and \autoref{eta_def}, we can propose a compact reformulation of the M\"obius transform of Eq. \ref{mob_trans} in the form of \autoref{emit}.}

\begin{lemma}[\textit{Compact reformulation of the M\"obius inversion formula}]\label{emit} 
	For any set $S \supseteq \supp{f}$, we have $\forall y \in P$:
	\begin{align}\label{general_f}
	{f(y) 
	= \sum_{\substack{X \in \suppartof{S}\\{y \in {\uparrow X}}\\
			{X \subseteq {\uparrow S}}
	}} \hspace{-0.3cm}g(X)~.~ \eta^{\phantom{\dagger}}_{S,\leq, P}(X, y)}
	\end{align}
	where $g(X) = g(x)$ for any $x \in X$. 
\end{lemma}	
	\begin{proof}
		See Appendix \ref{appendix:emit}.
	\end{proof}

{However, this reformulation is not a simplification and is thus of little use if our M\"obius function aggregate $\eta$ must be computed from the original M\"obius function $\mu$ as in \autoref{eta_def}. Fortunately, it is possible to compute $\eta$ recursively, independently from $\mu$, as shown in the following \autoref{eta_min_theorem}.}

\begin{lemma}[\textit{Recursive aggregation of M\"obius function images}]\label{eta_min_theorem}
	For any nonempty set of elements $S \subseteq P$ and for any part $X \in \suppartof{S}$,
		if every part $Z \in \suppartof{S}$ verifying $\lowerclosurein{S}{X} \subseteq \lowerclosurein{S}{Z}$ has a minimum, i.e. $|\min(Z)|=1$, then we have
		for any $y \in P$ where $\bigwedge X < y$: 
		\begin{align}\label{eta_min}
		{\eta^{\phantom{\dagger}}_{S,\leq, P}(X, y)
			= - \sum_{\substack{Z \in \suppartof{S}\\\bigwedge X < \bigwedge Z \leq y}} \eta^{\phantom{\dagger}}_{S,\leq, P}(Z, y),}
		\end{align}
		and $\eta^{\phantom{\dagger}}_{S,\leq, P}(X, y) = 1$ if $y = \bigwedge X$. 		
\end{lemma}
	\begin{proof}
		See Appendix \ref{appendix:eta_min_theorem}.
	\end{proof}

{In other words, for any part $X$ with a minimum $m$, if every part containing elements greater than $m$ has a minimum, then {${\eta}$ can be written in a recursive form} that only depends on itself. In fact, it is easy to see that $\eta$ is an extension of the M\"obius function $\mu$ of Eq. \ref{mobius_func} associated with the partially ordered set made of every minimum of part that is in the upper closure of $S$ (See \autoref{eta_fp}). So, the minimum of each of these parts constitutes one of the \textit{representatives} we were looking for in section \ref{intuition}, i.e. our focal points. The following section \ref{fp_section} will focus on them.}

{	
	Notice that we specifically target the parts that are in the upper closure of $S$ in order to avoid any unnecessary constraint on $\min(P)$. Indeed, the lowest parts contain minimal elements of $P$ and may not have a minimum if $P$ does not have one. However, if they are not in the upper closure of S, then their image through both $f$ and $g$ is 0, which means that they have no influence in Equations \ref{compact_zeta} and \ref{general_f} and so do not contribute to the definition of $f$ nor $g$. They are not used in Eq. \ref{eta_min} either since this recursion goes upward in $P$. On the other hand, if they are in the upper closure of $S$, then each of them necessarily has an element $s$ of $S$ as its minimum since there is no other part below it to contain $s$ and since $s$ is the least element of $P$ that is greater than or equal to $s$. Thus, we simply ignore the parts that are outside the upper closure of $S$.
}

\subsection{\textbf{Focal points and their implications}}\label{fp_section}

The purpose of \autoref{focal_origin} is to introduce the join-closure operator that will be used to formalize the notion of focal point. Then, \autoref{non_fp} will give the image through $f$ and $g$ of all non focal points. Finally, \autoref{eta_fp} will define the M\"obius function extension to use in \autoref{emitfp}, which contains what we called the \textit{Efficient M\"obius inversion formula}.

\begin{definition}[\textit{Join-closure}]\label{focal_origin}
	For any nonempty set of elements $S \subseteq P$, we note 
	$\supfpof{S}$ the smallest 
	join-closed subset of $P$ containing $S$, i.e.:
	\begin{empheq}{align*}
	\supfpof{S} = \left\lbrace \bigvee F ~/~ \emptyset \subset F \subseteq S \right\rbrace
	\end{empheq}
	The operator $^\vee \cdot:2^P\rightarrow2^P$ is thus a closure operator, i.e. for any sets $S, S' \subseteq P$, we have: 
	$S \subseteq \supfpof{S}{,}$ $\quad S \subseteq S' \Rightarrow {\supfpof{S}} \subseteq \supfpof{S'}$ \text{ and } $\supfpof{(\supfpof{S})} = \supfpof{S}$.
	This notion is illustrated in Fig. \ref{fig:focal_points}. 	
\end{definition}

{By definition of the supremum, i.e. the {least} upper bound, each element of \supfpof{S} is the {minimum} of a part $X \in \suppartof{S}$ verifying $\lowerclosurein{S}{X} \neq \emptyset$. Reciprocally, if a part $X \in \suppartof{S}$ verifying $\lowerclosurein{S}{X} \neq \emptyset$ has a minimum, then it is in \supfpof{S}. Therefore, \supfpof{S} is the set made of the minimum of each part from $\suppartof{S}$ in the upper closure of $S$. 
	In particular, we have $S \subseteq \supfpof{S}$. Yet, parts that do not contain any element of $S$ may not have a minimum. Thus, it may be necessary to check the existence of all these minima before anything. It is equivalent to checking that \supfpof{S} is an upper subsemilattice of $P$, i.e. a subset of $P$ for which the supremum (as defined in $P$) of every nonempty subset exists in it. For instance, if $P$ is itself an upper semilattice, then all these minima exist. 
}

{The elements of $\supfpof{\supp{f}}$ are what we call \textit{focal points}\footnote{This name stands for an analogy in the field of optics: a focal point is the point of the spatial domain ($P$) where an image is formed by the intersection of rays coming from a distant source ($\supp{f}$) passing through a lense (subset of $\supp{f}$).}. }
	
\begin{example}\label{ex_fp2}
	Let $(P, \leq)=(2^\Omega, \subseteq)$, where $\Omega = \{ a, b, c \}$. In this partially ordered set, the supremum operator $\vee$ is the union operator $\cup$.
	Let $m$ be a mass function such that $m(\Omega) = 0.1$, $m(\{ a, b \}) = 0.1$, $m(\{ b, c \}) = 0.2$ and $m(\{ a \}) = 0.6$. We have $\supp{m} = \{\Omega, \{ a, b \}, \{ b, c \}, \{ a \} \}$.
	It is easy to see that the union of any selection of support elements gives another support element. Therefore, we have $\supfp{m} = \supp{m}$. 
	
	It may help, for one that is familiar with DST, to notice that $\supfp{m} = \supp{m'}$, where $m' = \underset{s \in \supp{m}}{\osymbol{\cup}} m$ and $\osymbol{\cup}$ is the disjunctive fusion operator.
\end{example}

\begin{example}\label{ex_fp}
	Taking back \autoref{ex_fp2}, but looking at the dual closure operator $^\wedge\cdot$, we get in particular $\bigcap\{\{ a, b \}, \{ b, c \}\} = \textbf{\{ b \}}$ and $\bigcap\{\{ b \}, \{ a \}\} = \boldsymbol{\emptyset}$. We have $\inffp{m} = \{ \Omega, \{ a, b \}, \{ b, c \}, \{ a \}, \{ b \}, \emptyset \} = \supp{m} \cup \{ \{b\}, \emptyset \}$. This meet-closed subset of $P$ contains the focal points of $m$ in $(2^\Omega, \supseteq)$.
	
	It may help, for one that is familiar with DST, to notice that $\inffp{m} = \supp{m'}$, where $m' = \underset{s \in \supp{m}}{\osymbol{\cap}} m$ and $\osymbol{\cap}$ is the conjunctive fusion operator.
\end{example}
	
	{By definition of a level partition (\autoref{suppart_def}), we get \autoref{non_fp}, which determines the image through $f$ and $g$ of all elements of $P$ that are not focal points.}

\begin{property}[\textit{Images of non focal points}]\label{non_fp}
	For any upper subsemilattice $\supfpof{S}$ of $P$ such that $\supfpof{S} \supseteq \supp{f}$, and for any element $y \not\in \supfpof{S}$,
	we have $f(y) = 0$. Also, if $y \in {\uparrow \supfpof{S}}$, then $g(y) = g(s),$ where $y$ \textit{covers} $s$ in $\supfpof{S}$, i.e. $s$ is the maximum among the elements of $\supfpof{S}$ lower than $y$. Otherwise, $g(y)=0$.
\end{property}

\begin{example}
	Taking back $m$ from \autoref{ex_fp2} and the implicability function $b$ from \autoref{m_to_b}, we see that the elements from $2^\Omega$ that are not focal points are $\{a,c\}$, $\{c\}$, $\{b\}$ and $\emptyset$. So, we have $m(\{a,c\}) = m(\{c\}) = m(\{b\}) = m(\emptyset) = 0,\quad$ $b(\{a,c\}) = b(\{a\})$ and $b(\{c\}) = b(\{b\}) = b(\emptyset) = 0$.
\end{example}

\begin{example}\label{non_fp_ex_q}
	Taking back $m$ from \autoref{ex_fp} and the commonality function $q$ from \autoref{m_to_q}, we must look this time at the minimum among the elements of $\inffp{m}$ greater than some non focal point. We see that the elements from $2^\Omega$ that are not focal points are $\{a,c\}$ and $\{c\}$. So, we have $m(\{a,c\}) = m(\{c\}) = 0$, $q(\{a,c\}) = q(\Omega)$ and $q(\{c\}) = q(\{b,c\})$.
\end{example}

{Furthermore, thanks to \autoref{eta_min_theorem} and \autoref{focal_origin}, we can now define (\autoref{eta_fp}) the extension of the M\"obius function to be applied in our compact reformulation of the M\"obius transform.}

\begin{definition}[\textit{M\"obius function extension}]\label{eta_fp}
	For any nonempty set of elements $S \subseteq P$ such that \supfpof{S} is an upper subsemilattice of $P$, we define the extension $\eta^{\phantom{\dagger}}_{S,\leq, P}: \supfpof{S}\times P \rightarrow \mathbb{Z}$ of the M\"obius function $\mu^{\phantom{\dagger}}_{\supfpof{S},\leq}: \supfpof{S} \times \supfpof{S} \rightarrow \mathbb{Z}$ as follows:
	
	For any part $X \in \suppartof{S}$ such that $X \subseteq {\uparrow S}$ and for any $y \in P$ where $\bigwedge X < y$, 
	\begin{align*}
	\eta^{\phantom{\dagger}}_{S,\leq, P}\left({\bigwedge X}, y\right) = \eta^{\phantom{\dagger}}_{S,\leq, P}(X, y) = - \sum_{\substack{Z \in \suppartof{S}\\\bigwedge X < \bigwedge Z \leq y}} \eta^{\phantom{\dagger}}_{S,\leq, P}\left(\bigwedge Z, y\right),
	\end{align*}
	which is equivalent to stating that
	for any $(s,y) \in\supfpof{S}\times P$ where $s < y$,
	\begin{align}\label{eta_supfp}
	{\eta^{\phantom{\dagger}}_{S,\leq, P}(s, y)
		= - \sum_{\substack{p \in \supfpof{S}\\s < p \leq y}} \eta^{\phantom{\dagger}}_{S,\leq, P}(p, y),}
	\end{align}
	with $\eta^{\phantom{\dagger}}_{S,\leq, P}(s, s) = 1$.	
\end{definition}

{The final part of this section consists in proposing our so-called \textit{Efficient M\"obius inversion formula} in the form of \autoref{emitfp}, which exploits the compact reformulation of \autoref{emit}, the focal points of \autoref{focal_origin} and the extended M\"obius function defined in \autoref{eta_fp}.}

\begin{theorem}[\textit{Efficient M\"obius inversion formula}]\label{emitfp} 
	For any upper subsemilattice $\supfpof{S}$ of $P$ such that $\supfpof{S} \supseteq \supp{f}$, we have $\forall y \in P$,
	\begin{align}\label{general_f_supfp}
	{f(y) 
		= \sum_{\substack{s \in \supfpof{S}\\s \leq y}} \hspace{-0.cm}g(s)~.~ \eta^{\phantom{\dagger}}_{S,\leq, P}(s, y)}
	\end{align}
	
\end{theorem}	
\begin{proof}
	Eq. \ref{general_f_supfp} is a simple reformulation of Eq. \ref{general_f} from \autoref{emit} with the focal points of \autoref{focal_origin} and the extended M\"obius function $\eta$ of \autoref{eta_fp} (given \autoref{eta_min_theorem}), combined with the fact that $\supfpof{(\supfpof{S})} = \supfpof{S}$, which also means that $\suppartof{\supfpof{S}} = \suppartof{S}$. 
\end{proof}

{Notice that {${S}$ in fact need not contain ${\supp{f}}$}, as long as $\supfpof{S}$ does. So, for example, $\supfpof{S}$ can be a sublattice of $P$ verifying $\supfpof{S} \supseteq \supp{f}$, with $S = I \cup \{\bigwedge I\}$, where $I$ is the set containing the join-irreducible elements of $\supfpof{S}$, as is the case with the \textit{lattice support} from \cite{me}, where $S$ may not contain $\supp{f}$. Also, note that $\suppartof{\supfpof{S}} = \suppartof{S}$ since $\supfpof{(\supfpof{S})} = \supfpof{S}$, which means that all computations can be made only based on the upper semillatice $\supfpof{S}$, without actually having to determine any set $S$.}

\begin{example}
	Taking back \autoref{b_to_m}, we get that for any focal point $s\in\supfpof{\supp{m}}$, the M\"obius transform $m$ of $b$ in $(2^\Omega, \subseteq)$ is the M\"obius transform of $b$ in $(\supfpof{\supp{m}}, \subseteq)$, i.e. noting $S = \supp{m}$, we have
	\begin{align*}
	\forall y \in \supfpof{S},\quad m(y) = \sum_{\substack{s \in \supfpof{S}\\s \subseteq y}} b(s)~.~ \mu^{\phantom{\dagger}}_{\supfpof{S}, \subseteq}(s, y) 
	\end{align*}
\end{example}

\begin{example}
	Similarly, taking back \autoref{q_to_m}, we get that for any focal point $s\in\inffpof{\supp{m}}$, the M\"obius transform $m$ of $q$ in $(2^\Omega, \supseteq)$ is the M\"obius transform of $q$ in $(\inffpof{\supp{m}}, \supseteq)$, i.e. noting $S = \supp{m}$, we have
	\begin{align*}
	\forall y \in \inffpof{S},\quad m(y) = \sum_{\substack{s \in \inffpof{S}\\s \supseteq y}} q(s)~.~ \mu^{\phantom{\dagger}}_{\inffpof{S}, \supseteq}(s, y) 
	\end{align*}
\end{example}

\pgfplotsset{
	ylabel style={rotate=-90},
}

\subsection{\textbf{Ways to compute focal points}}\label{fp_computation}

This section focuses on methods allowing one to compute focal points in an efficient way.  
\autoref{dotF_computation} describes a direct scheme for computing the join-closure $\supfpof{S}$ of any set $S\subseteq P$, while \autoref{fp_from_g} indicates how to find $\supfp{f}$ based on $\mathcal{G}$ alone.

\begin{property}[\textit{Computing the join-closure of $S$ directly}]\label{dotF_computation}
	Every element in $\supfpof{S}$ can be described as either $y$, where $y \in S$, or $s \vee y$, where $s \in \supfpof{S}$. Doing so, all focal points can be found through a double loop: the outer one iterating through $S$ and the inner one dynamically iterating through already found focal points (starting with $S$).
	Therefore, all elements of $\supfpof{S}$ can be found in $O(|\supfpof{S}|.|S|)$.
	It is even possible to further optimize since for any $x,y \in P$, if $x \leq y$ or $x \geq y$, then $x \vee y = y$ or $x \vee y = x$, which means in our case that it is useless to compute the supremum of two elements if there exists an order between them.
\end{property}

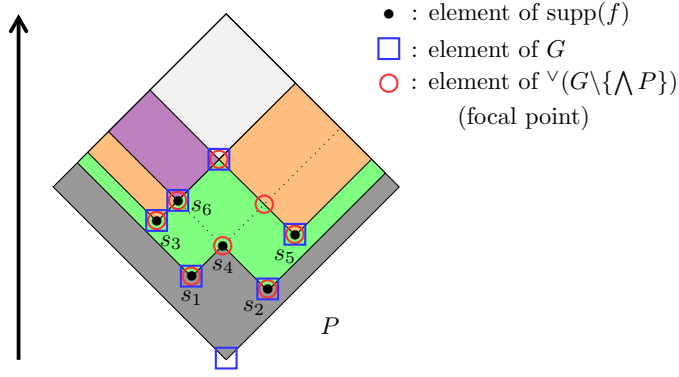
\begin{figure}[t]
	\centering
	\hspace{2cm}
	\begin{tikzpicture}[scale=0.92, every node/.style={transform shape},
	node/.style={draw, dot,minimum size=0.2cm, inner sep=0pt},
	det/.style={draw, diamond,minimum size=1.1cm, inner sep=0pt},
	rect/.style={draw, rectangle,minimum size=1.1cm, inner sep=2pt}
	]
	
	\filldraw[fill=white!20!gray, draw=black] (0,0) -- (2.5, 2.5) -- (0,5) -- (-2.5,2.5) -- (0,0);
	\filldraw[fill=white!50!green, draw=black] (-0.5,1.2) -- (-2.15, 2.85) -- (0,5) -- (2.3, 2.7) -- (0.6, 1) -- (-0.05, 1.65) -- (-0.5,1.2);
	
	
	\filldraw[fill=white!50!orange,] (-1,2) -- (-2, 3) -- (0,5) -- (2.1, 2.9) -- (1,1.8) -- (-0.105, 2.9) -- (-1,2);
	\draw[dotted] (-1.7, 3.3) -- (-0.05, 1.65) -- (1.65, 3.35);
	\filldraw[fill=white!50!violet] (-0.7, 2.3) -- (-1.7, 3.3) -- (0, 5) -- (1, 4) -- (-0.7, 2.3);
	\filldraw[
	fill=white!90!gray, 
	draw=black] (-0.105, 2.9) -- (-1.1, 3.9) -- (0,5) -- (1,4) -- (-0.105, 2.9);
	\node (P) at (1.5, 0.5) {$P$};	
	\node (s1) at (-0.5, 0.9) {$s_1$};
	\node (s2) at (0.4, 0.75) {$s_2$};
	\node (s3) at (-0.8, 1.7) {$s_3$};
	\node (s4) at (-0.045, 1.35) {$s_4$};
	\node (s5) at (0.85, 1.5) {$s_5$};
	\node (s6) at (-0.35, 2.2) {$s_6$};
	\node (a) at (-0.4974, 1.1975) {$\bullet$};
	\node (b) at (-1, 2) {$\bullet$};
	\node (c) at (1, 1.8) {$\bullet$};
	\node (d) at (-0.695, 2.285) {$\bullet$};
	\node (e) at (0.605, 1.015) {$\bullet$};
	\node (f) at (-0.045, 1.64) {$\bullet$};
	\draw[line width=0.05cm,-{angle 60[length=100mm, width=2000mm]}] (-3,0) -- (-3., 5);
	\node[draw,thick,white!20!red,circle,minimum size=0.25cm,inner sep=0pt] (afp) at (-0.505, 1.21) {};
	\node[draw,thick,white!20!blue,rectangle,minimum size=0.3cm,inner sep=0pt] (afp) at (-0.505, 1.21) {};
	\node[draw,thick,white!20!red,circle,minimum size=0.25cm,inner sep=0pt] (bfp) at (-1.005, 2.015) {};
	\node[draw,thick,white!20!blue,rectangle,minimum size=0.3cm,inner sep=0pt] (afp) at (-1.005, 2.015) {};
	\node[draw,thick,white!20!red,circle,minimum size=0.25cm,inner sep=0pt] (cfp) at (0.9925, 1.815) {};
	\node[draw,thick,white!20!blue,rectangle,minimum size=0.3cm,inner sep=0pt] (afp) at (0.9925, 1.815) {};
	\node[draw,thick,white!20!red,circle,minimum size=0.25cm,inner sep=0pt] (acfp) at (0.55, 2.25) {};
	\node[draw,thick,white!20!red,circle,minimum size=0.25cm,inner sep=0pt] (abfp) at (-0.105, 2.9) {};
	\node[draw,thick,white!20!blue,rectangle,minimum size=0.3cm,inner sep=0pt] (afp) at (-0.105, 2.9) {};
	\node[draw,thick,white!20!red,circle,minimum size=0.25cm,inner sep=0pt] (abfp) at (0.6, 1.025) {};
	\node[draw,thick,white!20!blue,rectangle,minimum size=0.3cm,inner sep=0pt] (afp) at (0.6, 1.025) {};
	\node[draw,thick,white!20!red,circle,minimum size=0.25cm,inner sep=0pt] (abfp) at (-0.7, 2.305) {};
	\node[draw,thick,white!20!blue,rectangle,minimum size=0.3cm,inner sep=0pt] (afp) at (-0.7, 2.305) {};
	\node[draw,thick,white!20!red,circle,minimum size=0.25cm,inner sep=0pt] (abfp) at (-0.05, 1.65) {};
	\node[draw,thick,white!20!blue,rectangle,minimum size=0.3cm,inner sep=0pt] (afp) at (0, 0.025) {};
	\node (l1) at (4.05,5) {$\bullet~$ : element of $\supp{f}$};
	\node[draw,thick,white!20!red,circle,minimum size=0.25cm,inner sep=0pt] (l2) at (2.35,4) {};
	\node (l2desc) at (4.6,4) {: element of $\supfpof{(G\backslash \{ \bigwedge P \})}$};
	\node (l2descsub) at (4.3,3.5) {(focal point)};
	\node[draw,thick,white!20!blue,rectangle,minimum size=0.3cm,inner sep=0pt] (afp) at (2.35,4.5) {};
	\node (l2desc) at (3.785,4.5) {: element of $G$};
	\end{tikzpicture}
	\caption{\small
		Example of image partition $\mathcal{G}$ from Fig. \ref{fig:image_partition}. Each color represents an image through $g$. Suppose that $g$ is the zeta transform of $f$ in $(P, \leq)$ and that the image through $g$ of the gray part is 0. Then, this partition is the consequence of the support of $f$ displayed in the figure with the following conditions: $f(s_2) = f(s_1)$, $f(s_4) = -f(s_1)$ and $f(s_3) = f(s_5)$. In this case, the image through $g$ of the elements in the green part is $f(s_1)$, the one in the orange part is $f(s_1) + f(s_3)$, the one in the violet part is $f(s_1)+f(s_3)+f(s_6)$ and the one in the white part is $f(s_1)+2.f(s_3)+f(s_6)$. Notice that not all elements from $\supp{f}$ are minimal elements of a part. However, they are all contained in the join-closure of $G\backslash \{ \bigwedge P \}$, noted $\supfpof{(G\backslash \{ \bigwedge P \})}$, where $G = \bigcup_{X \in \mathcal{G}} \min(X)$. In fact, we even have $\supfpof{(G\backslash \{ \bigwedge P \})} = \supfp{f}$.
	}\label{fig:partition_link}
\end{figure}

{
This \autoref{dotF_computation} can be used to compute focal points directly from $\supp{f}$, but $\supp{f}$ is not always known beforehand. One may want to find all focal points from $g$ alone. The problem is that for any two elements $x, y \in P$, $g(x) \neq g(y) \Rightarrow \lowerclosurein{S}{x} \neq \lowerclosurein{S}{y}$, but the converse is not true. Therefore, some elements from $\supfp{f}$ may not be directly apparent in the compact definition of $g$ through its image partition $\mathcal{G}$. See Fig. \ref{fig:partition_link}. The following \autoref{fp_from_g} explains how focal points can be found from $\mathcal{G}$ alone, in spite of this fact.}

\begin{theorem}[\textit{Finding $\supfp{f}$ from $\mathcal{G}$}]\label{fp_from_g}
	{Let $Y$ be the set made of every supremum $x \vee a$, where $x \in G\backslash M$ and $a \in M$,
		where $M = \min(P) \cap \overline{\supp{g}}$ and $G = \bigcup_{X \in \mathcal{G}} \min(X)$.
	If $\supfpof{G}$ is an upper subsemilattice of $P$ (e.g. if $P$ is itself an upper semilattice), then we have $\supfpof{\supp{f}} \subseteq \supfpof{(Y\cup G\backslash M)} \subseteq \supfpof{G}$.}
\end{theorem}
\begin{proof}
	See Appendix \ref{appendix:fp_from_g}.
\end{proof}

Consequently, one can find all focal points of $f$ and $g$, from either $\supp{f}$ or the minimal elements of $\mathcal{G}$, either using a variation of the procedure described in \autoref{dotF_computation} or by building a sublattice $L$ of $P$ as done in Proposition 2 of \cite{me}. This latter method, of complexity $O(n.|L|)$, potentially generates more elements but features a better worst-case complexity $O(n.|P|)$, where $n$ is the number of join-irreducible elements of $L$. In addition, this lattice $L$ contains both $\supfp{f}$ and $\inffp{f}$.

It is also worth noting that we do not even have to compute the join-closure of $G$ if $f$ is nonnegative, since there can be no compensation of images through $f$. This means that the image through $g$ of a focal point cannot be equal to the one of an element it covers (See Appendix \ref{appendix:fp_from_g}). Hence \autoref{fp_from_g_f_pos}.

\begin{property}[\textit{Finding $\supfp{f}$ from $\mathcal{G}$ when $f$ is nonnegative}]\label{fp_from_g_f_pos}
	If $\supfpof{\supp{f}}$ is an upper subsemilattice of $P$ (e.g. if $P$ is itself an upper semilattice) and if $f$ is nonnegative, then we have $\supfpof{\supp{f}} \subseteq G\backslash M$, where $M = \min(P) \cap \overline{\supp{g}}$ and $G = \bigcup_{X \in \mathcal{G}} \min(X)$.
\end{property}

\begin{example}
	The mass function $m$
	is required to be nonnegative. Therefore, no matter the partition defining the implicability function $b$ and the commonality function $q$,
	the partition for $b$ always contains its focal points among the minimal elements of its parts, and the partition for $q$ (the dual of $b$) always contains its focal points among the maximal elements of its parts (provided that $q$ and $b$ indeed correspond to mass functions).
	However, the disjunctive weight function $v$ from \autoref{b_to_v} and the conjunctive weight function $w$ from \autoref{q_to_w}, which are multiplicative M\"obius transforms in DST, are allowed to have values below 1. Thus, they do not satisfy the multiplicative equivalent of \autoref{fp_from_g_f_pos}. 
\end{example}


\newcommand{\lowerclosurefp}[1]{\sloppy\text{${^{\supfp{f},\downarrow} #1}$}}
\newcommand{\lowerclosurefpin}[2]{\sloppy\text{${^{\supfpof{#1},\downarrow} #2}$}}
\newcommand{\upperclosurefp}[1]{\sloppy\text{${^{\inffp{f},\uparrow} #1}$}}
\newcommand{\upperclosurefpin}[2]{\sloppy\text{${^{\inffpof{#1},\uparrow} #2}$}}

\newcommand{\proxy}[3]{\sloppy\text{proxy$(#1,#2,#3)$}}
\newcommand{\proxydef}[3]{\sloppy\text{proxy$(#1,#2,#3) = \{ p \in {^{#1,#2}#3}~/~{^{#1,#2}p} = {^{#1,#2}#3}\backslash \{#3\} \}$}}


\subsection{\textbf{Focal points for both additive and multiplicative M\"obius transforms}}\label{link_section}

\newcommand{\lowerclosureh}[1]{\sloppy\text{${^{\supp{h-1},\downarrow} #1}$}}
\newcommand{\upperclosureh}[1]{\sloppy\text{${^{\supp{h-1},\uparrow} #1}$}}
\newcommand{\lowerclosurefph}[1]{\sloppy\text{${^{\supfp{h-1},\downarrow} #1}$}}
\newcommand{\upperclosurefph}[1]{\sloppy\text{${^{\inffp{h-1},\uparrow} #1}$}}

In this section, we will see which place takes the focal points of two functions $f:P \rightarrow \mathbb{R}$ and $h:P \rightarrow \mathbb{R}^*$ when they are linked by the following equation:
\begin{align}\label{link}
\forall y \in P,\quad&g(y) = \sum_{x \leq y} f(x) = \prod_{x \leq y} h(x)
\end{align}
We will not consider the case where $h$ is allowed to have null images since they make impossible the inversion of products. Indeed, any element of $P$ that is greater than an element associated with a null value through $h$ is guaranteed to get a null image through the zeta transform $g$, no matter what image it has through $h$. Therefore, retrieving their image through $h$ from $g$ becomes impossible. The following \autoref{supp_zero} reflects this constraint.

\begin{property}[]\label{supp_zero}
	For any minimal element $y \in \min(P)$, Eq. \ref{link} gives us $g(y) = f(y) = h(y)$. In particular, if $y \not\in \supp{f}$, then $g(y)=f(y)=h(y)=0$, which we forbid.
	Thus, we have $\min(P) \subseteq \supp{f}$.
\end{property}

\begin{example}\label{bmv}
	Taking back the implicability function $b$ from \autoref{m_to_b} and the disjunctive weight function $v$ from \autoref{b_to_v}, we have:
	\begin{align*}
	\forall y \in 2^\Omega,\quad&b(y) = \sum_{x \subseteq y} m(x) = \prod_{x \subseteq y} v(x)^{-1},
	\end{align*}
	which implies that $m(\emptyset) = v(\emptyset)^{-1}$ and so $m(\emptyset) \neq 0$, which is in accordance with \autoref{supp_zero}.
\end{example}
\begin{example}\label{qmw}
	Taking back the commonality function $q$ from \autoref{m_to_q} and the conjunctive weight function $w$ from \autoref{q_to_w}, we have:
	\begin{align*}
	\forall y \in 2^\Omega,\quad&q(y) = \sum_{x \supseteq y} m(x) = \prod_{x \supseteq y} w(x)^{-1},
	\end{align*}
	which implies that $m(\Omega) = w(\Omega)^{-1}$ and so $m(\Omega) \neq 0$, which is in accordance with \autoref{supp_zero}.
\end{example}

From this \autoref{supp_zero} and \autoref{fp_from_g}, we can link the focal points of $f$ and $h$, and thus their respective \textit{support} elements, in \autoref{emi_link_fp_f_full}.

\newcommand{\supparth}{\sloppy\text{$P/(\supp{h-1}, \leq)$}}
\newcommand{\suppartdualh}{\sloppy\text{$P/(\supp{h-1}, \geq)$}}

\begin{corollary}[\textit{Link between $\supfp{f}$ and $\supfp{h-1}$}]\label{emi_link_fp_f_full}
	If {either} $\supfpof{(\supp{h-1} \cup \min(P))}$ {or} $\supfpof{\supp{f}}$ is an upper subsemilattice of $P$,  
	then we have 
	$${\supfpof{\supp{f}} = \supfpof{(\supp{h-1}\cup \min(P))}}$$ 
\end{corollary}
	\begin{proof}
		See Appendix \ref{appendix:emi_link_fp_f_full}.
	\end{proof}

In particular, $\supfp{f} = \{\bigwedge P\} \cup \supfp{h-1}$, if $P$ has a minimum. 

\begin{example}
	Taking back the mass function $m$, the disjunctive weight function $v$ and the conjunctive weight function $w$ from \autoref{bmv} and \autoref{qmw}, we get $\supfpof{\supp{m}} = \{ \emptyset \} \cup \supfpof{\supp{v-1}}$ and $\inffpof{\supp{m}} = \{ \Omega \} \cup \inffpof{\supp{w-1}}$.
\end{example}

Finally, \autoref{ablation_study_generalized} proposes formulas to track the information flow from $h$ to $g$ and $f$, i.e. the effects of changing one element in $\supp{h-1}$ on $f$ and $g$. This will be used in the application of section \ref{dst_ablation}. The reversed flow, from $f$ to $h$, is not displayed here as it does not simplify well.

\begin{theorem}[\textit{Information flow from $h$ to $g$ and $f$}]\label{ablation_study_generalized}
	Let $h'$ be equal to $h$ everywhere, except for the image of some $x \in P$. Also, let $f'$ and $g'$ be the functions corresponding to $h'$ so that they satisfy Eq. \ref{link} in place of respectively $f$ and $g$. Then we have:
	\begin{empheq}{align}\label{influence_w_g}
	\forall y \in P,\quad g'(y) = 
	\begin{cases}
	g(y)~.~ \frac{h'(x)}{h(x)} &\text{if $x \leq y$}\\
	g(y) &\text{otherwise}
	\end{cases}
	\end{empheq}
	and
	for any upper subsemilattice $\supfpof{S}$ of $P$ such that $\supfpof{S} \supseteq \supp{f} \cup \{x\}$,
	\begin{empheq}{align*}
	\forall y \in P,\quad f'(y) = \begin{cases}
	0	&\text{if $y \not\in \supfpof{S}$}\\
	f(y) + \left[\frac{h'(x)}{h(x)}-1\right].f_{\uparrow x}(y) &\text{if $x \leq y$}\\
	f(y)	&\text{otherwise}
	\end{cases}
	\end{empheq}
	where $f_{\uparrow x} : {\uparrow x} \rightarrow \mathbb{R}$ is the M\"obius transform of $g$ in $({\uparrow x}, \leq)$, i.e. $\forall y \in {\uparrow x}$, 
	$$f_{\uparrow x}(y) 
	= \sum_{\substack{s \in {\uparrow x}\\s \leq y}} g(s)~.~\mu^{\phantom{\dagger}}_{\uparrow x, \leq}(s, y)
	$$
\end{theorem}	
\begin{proof}
See Appendix \ref{appendix:ablation_study_generalized}.
\end{proof}

{In particular, notice that if ${x \in \supfp{f}}$, then ${\supfp{f} \supseteq \supp{f} \cup \{x\}}$}.

\subsection{\textbf{Discussions}}\label{discussions}

Several remarks can be made regarding $g$ as input. 
Firstly, we only consider a compact definition of $g$ in this article because the point is that it is possible to compute its M\"obius transform even in a tremendously vast partially ordered set by avoiding to consider all elements of $P$. If $g$ is defined by its image on every element of $P$, then the usual M\"obius transform should be employed, as the search for focal points would require to check all elements of $P$ at least once. This is all the more relevant since there are some algorithms like the FMT \cite{FMT} that operate in $O(n.|P|)$, where $P$ is here a lattice and $n$ is the number of its join-irreducible elements, where we usually have $n \ll |P|$.

Secondly, if $P$ is downward infinite, then there may be parts in $\mathcal{G}$ that do not have any minimal element. Since our method may need these minimal elements in order to find some potentially \textit{hidden} support elements if $f$ can have negative values, one might have to add a surrogate element for each downward infinite horizon in $P$. For example, if $P$ contains an element that is greater than two infinite chains, then one should add two surrogate elements symbolizing downward infinity, one for each chain. Of course, this can only be possible if there is a finite number of downward infinite horizons. 
Finally, $|\supp{f}|$ must be finite and $\mathcal{G}$ must have a finite number of parts. Otherwise, focal points cannot be determined. 

\subsection{\textbf{From theory to practice}}\label{theory_practice}

In practice, one could wonder how these formulas can be exploited. Actually, even though it is possible to do so, the M\"obius function $\mu$ (or our extension $\eta$) is usually not evaluated when computing the M\"obius transform of a function. It mostly serves a theoretical purpose. Instead, it is enough to consider the zeta transform $g$ of a function $f$ and to simply rewrite it in a recursive way to express $f$ in terms of $g$ and $f$:
\begin{align}\label{zeta_inverse}
\forall y \in P,\quad g(y) = \sum_{\substack{x \in S\\x \leq y}} f(x)	\quad\Leftrightarrow\quad		f(y) = g(y) - \sum_{\substack{x \in S\\x < y}} f(x)
\end{align}
where $S \supseteq \supp{f}$.
Then, there are two main ways to use these formulas: (i) naively, summing all terms for each element $y$, or (ii) efficiently, reusing partial sums common to multiple elements $y$. With (i), values of $g$ can be computed independently, while (ii) requires the computation of the image through $g$ of all elements of $P$. But, if one needs the value of $g$ on every element $y$ (i.e. the complete definition of $g$), whether it is because $|\supp{f}|$ is close to $|P|$ or otherwise, then (ii) is much more efficient. The optimal method achieving (ii) for $P=2^\Omega$ is the \textit{Fast M\"obius Transform} (FMT) \cite{FMT}, which has a time complexity in $O(N.2^N)$, where $N$ is the size of the frame of discernment resulting from the best lossless coarsening\footnote{A lossless coarsened frame of discernment $\Omega'$ is a partition of the original set $\Omega$, subject to this coarsening, such that every support element of the considered mass function defined on $2^\Omega$ can be mapped into $2^{\Omega'}$. The best lossless coarsening results in the smallest $\Omega'$ possible (see \cite{wilson2000}).} of $\Omega$ regarding $\supp{f}$.

{Our contribution} to this is the proof that we do not necessarily need the value of ${g}$ on all elements of ${P}$ to define ${g}$ and to define ${f}$ from ${g}$ (without knowledge about $\supp{f}$). Our {focal points} constitute a subset of $P$ that can be substantially smaller and is {necessary and sufficient to define all zeta and M\"obius transforms}. 
As seen in \autoref{emitfp}, for any element of $\supfpof{S}$ the M\"obius transform of $g$ in $(P, \leq)$ is the M\"obius transform of $g$ in $(\supfpof{S}, \leq)$. This means that we can work in the domain of our focal points instead of $P$ and obtain the exact same results. This finding is mostly useful for (ii) as it exploits the structure of the domain. For this, we proposed variants of the FMT, called \textit{Efficient M\"obius Transformations} ({EMT}) \cite{me}, which {work in any distributive lattice ${L}$} and exploit the structure of its subsemilattices. {Taking ${L=2^\Omega}$, the complexity of the EMT is always lower than ${O(N.2^N)}$ and can be even lower than ${O(N^2)}$}, e.g. if $\supp{f}$ is a chain.

When using (i), only support elements are necessary in computations, but $g$ is only completely defined when its image on all focal points is given (See \autoref{ex_emt_1}), which is obvious considering its image partition $\mathcal{G}$.
In addition, it is important to mention that changing even only one image through $g$ of an element from $\supp{f}$ may add or remove elements from $\supp{f}$, but they will always be in $\supfp{f}$. Thus, manipulating $g$ only on the elements of $\supp{f}$ is highly unreliable.
On the contrary, focal points can serve as data structure to completely define $g$ and can be directly found from any compact definition of $g$, such as one by intervals.

Furthermore, $\supp{f}$ may not be known before actually computing $f$ from $g$. In this case, it is impossible to compute $f$ with only elements from $\supp{f}$. However, it is always possible to find the focal points of $f$ from $g$. 

\noindent See \autoref{ex_emt_2}.
\begin{example}[\textit{Computing $q$ from $m$}]\label{ex_emt_1}
	Let us take back \autoref{ex_fp}, i.e. $\Omega = \{ a, b, c \}$ and $m$ is a mass function such that $m(\Omega) = 0.1$, $m(\{ a, b \}) = 0.1$, $m(\{ b, c \}) = 0.2$ and $m(\{ a \}) = 0.6$. We have $\supp{m} = \{\Omega, \{ a, b \}, \{ b, c \}, \{ a \} \}$ and $\inffp{m} = \supp{m} \cup \{ \{b\}, \emptyset \}$.
	From $m$, we get its commonality function $q$ based on Eq. \ref{compact_zeta} with $\supp{m}$ on its focal poins $\inffp{m}$:
	{	\small\begin{itemize}
			\item $q(\Omega) = m(\Omega) = \boldsymbol{0.1}$
			\item $q(\{a,b \}) = m(\{a,b\}) + m(\Omega) = \boldsymbol{0.2}$
			\item $q(\{b,c \}) = m(\{b,c\}) + m(\Omega) = \boldsymbol{0.3}$
			\item $q(\{a \}) = m(\{a\}) + m(\{a,b \}) + m(\Omega) = \boldsymbol{0.8} $
			\item $q(\{b \}) = m(\{a,b \}) + m(\{b,c \}) + m(\Omega) = \boldsymbol{0.4}$
			\item $q(\emptyset ) = m(\{a \}) + m(\{a,b \}) + m(\{b,c \}) +  m(\Omega) = \boldsymbol{1} $
	\end{itemize}}
	As already shown in \autoref{non_fp_ex_q}, the rest of $2^\Omega$, which does not contain any focal point, is defined by \autoref{non_fp}: $q(\{a,c\}) = q(\Omega)$ and $q(\{c\}) = q(\{b,c\})$.
\end{example}
\begin{example}[\textit{Computing $w$ from $m$}]\label{ex_emt_2}
	Now, from \autoref{ex_emt_1}, suppose that we want to compute $w$. From \autoref{emi_link_fp_f_full}, we know that $\{\Omega\} \cup \inffp{w-1} = \inffp{m}$. In addition, we have $m(\Omega) = w(\Omega)^{-1} = 0.1 \neq 1$, which means that $\Omega \in \supp{w-1}$ and so $\inffp{w-1} = \inffp{m}$. 
	Then, adapting Eq. \ref{zeta_inverse} to the multiplicative form, we get:
	\begin{align*}
	\forall y \in 2^\Omega,\quad w(y) = \begin{cases}
	1 &\text{if $y \not\in \inffp{m}$}\\
	q(y)^{-1}.\displaystyle\prod_{\substack{s \in \inffp{m}\\s \supset y}} w(s)^{-1} &\text{otherwise}
	\end{cases} 
	\end{align*}
	This result and its dual for the disjunctive weight function $v$ were already the conclusion of one of our previous papers \cite{me_gretsi}.
	So, from this, we get the conjunctive weight function $w$:
	{	\small\begin{itemize}
			\item $w(\Omega) = q(\Omega)^{-1} = \boldsymbol{10}$
			\item $w(\{a,b \}) = \left[q(\{a,b\}) ~.~ w(\Omega) \right]^{-1} = \boldsymbol{0.5}$
			\item $w(\{b,c \}) = \left[q(\{b,c\}) ~.~ w(\Omega) \right]^{-1} = \boldsymbol{\frac{1}{3}}$
			\item $w(\{a \}) = \left[q(\{a\}) ~.~ w(\Omega) ~.~ w(\{a,b \}) \right]^{-1} = \boldsymbol{0.25}$
			\item $w(\{b \}) = \left[q(\{b\}) ~.~ w(\Omega) ~.~ w(\{a,b \}) ~.~ w(\{b,c \}) \right]^{-1} = \boldsymbol{1.5}$
			\item $w(\emptyset ) = \left[q(\emptyset) ~.~ w(\Omega) ~.~ w(\{a,b \}) ~.~ w(\{b,c \}) ~.~ w(\{a \}) ~.~ w(\{b \}) \right]^{-1} = \boldsymbol{1.6}$
	\end{itemize}}
	\noindent Since all these images are different from 1, we get here\footnote{From experience, all focal points of $m$ are most of the time also elements of the support of $w-1$ or $v-1$, to the point where we in fact never witnessed a case in which this was not true.} that $\inffp{m} = \supp{w-1}$.
\end{example}

\section{\textbf{Implications for Dempster-Shafer Theory}}\label{applications}

In DST, we work with $P = 2^\Omega$, which is a lattice, i.e. a set of which every nonempty subset has both a supremum and an infimum. Doing so, all focal points always exist for both the relations $\subseteq$ and $\supseteq$. We will see in section \ref{dst_emt} how our \textit{Efficient} M\"obius inversion formula impacts almost all representations of DST and how to fuse belief sources using focal points. In section \ref{dst_generalized_weights}, we will propose a generalization of the conjunctive decomposition of evidence to benefit from fusion rules such as the Cautious one \cite{disj_dec_cautious_bold}, even when the considered mass function is dogmatic. Finally, section \ref{dst_ablation} will provide formulas to study the impact of each decomposition  weight on the corresponding mass function.

\subsection{\textbf{Efficient representations in Dempster-Shafer Theory}}\label{dst_emt}

In DST, the mass function $m$ is central. It is considered as a generalization of the discrete Bayesian probability distribution. It is defined within the bounds of two constraints \cite{smets94}: one is that $m$ is nonnegative, and the other is 
\begin{align}\label{constraint_m_alone}
\quad\displaystyle\sum_{y\in 2^\Omega} m(y) = 1.
\end{align}

However, other representations are often used to analyse or fuse mass functions. With the exception of the \textit{pignistic probability} representation, all of them are linked to the zeta and M\"obius transforms. We already introduced them in our examples throughout this article: 

\noindent the implicability\footnote{This representation alone is linked to two other ones: the belief function $\textit{Bel}$, which is equal to $b$ when setting $m(\emptyset) = 0$, and the plausibility function $\textit{Pl}(y)= 1-b(\overline{y})$, $\forall y\in 2^\Omega$.} function $b$ from \autoref{m_to_b}, the commonality function $q$ from \autoref{m_to_q}, the disjunctive weight function $v$ from \autoref{b_to_v} and the conjunctive weight function $w$ from \autoref{q_to_w}. 

In addition, \autoref{ex_dempster} displays a classic use case demonstrating how the fusion of two functions of same type in DST can be performed with focal points.

\begin{example}[\textit{Efficient combination with Dempster's fusion rule}]\label{ex_dempster}

Dempster's combination rule $\osymbol{+}$ is defined as the normalized conjunctive rule $\osymbol{\cap}$, i.e. $\forall y \in 2^\Omega ~/~ y \neq \emptyset$:

\begin{align}
(m_{1} \osymbol{+} m_{2})(y) &= \frac{1}{K} (m_{1} \osymbol{\cap} m_{2})(y)\nonumber\\
&= \frac{1}{K} \sum_{\substack{s_1 \cap s_2 = y\\s_1 \in \supp{m_1}\\s_2 \in \supp{m_2}}} m_1(s_1)~.~ m_2(s_2)\label{inter_conj}\\
&= \frac{1}{K} \sum_{\substack{x \supseteq y}} q_1(x) . q_2(x) . {\mu^{\phantom{\dagger}}_{2^\Omega, \supseteq}(x, y)}\label{comm_conj_mu}
\end{align}
where $K = 1 - (m_{1} \osymbol{\cap} m_{2})(\emptyset)$. Eq. \ref{inter_conj} can be used by an evidence-based algorithm in $O(|\supp{m_1}|~.~|\supp{m_2}|)$, observing that each support element in $\supp{m_{12}}$ of the combined mass function $m_{12}$ is defined as the intersection $s_1 \cap s_2$ of a pair of support elements, where $s_1 \in \supp{m_1}$ and $s_2 \in \supp{m_2}$. Alternatively, Eq. \ref{comm_conj_mu} can be used by powerset-based algorithms such as the FMT (this application has been tackled in \cite{FMT}) in $O(N.2^N)$, where $N$ is here the size of the frame of discernment resulting from the best lossless coarsening of $\Omega$ regarding $\supp{m_1}\cup\supp{m_2}$. It consists in separately computing $q_1$ and $q_2$ with the FMT and then multiplying them element-wise to get $q_{12}$, before using again the FMT to compute $m_{12}$ from $q_{12}$.
This second approach is useful when almost all images of $m_{12}$ are required, e.g. when $|\supp{m_1}|~.~|\supp{m_2}|$ is of significantly higher magnitude than $N.2^N$.

We can also reformulate Eq. \ref{comm_conj_mu} with focal points in the light of \autoref{emitfp} and obtain a hybrid approach:
\begin{align}\label{comm_conj}
(m_{1} \osymbol{+} m_{2})(y) &= 
\begin{cases}
\displaystyle\frac{1}{K} \sum_{\substack{s \in {\inffpof{S}}\\s \supseteq y}} q_1(s) . q_2(s) . {\mu^{\phantom{\dagger}}_{\inffpof{S}, \supseteq}(s, y)}	& \text{if $y \in \inffpof{S}$}\\
0 & \text{otherwise}
\end{cases}
\end{align}
where $\inffpof{S} \supseteq \supp{m_{12}}$ and $P = 2^\Omega$.
Eq. \ref{comm_conj} can be exploited in less than 
$O(N.2^N)$ with the EMT using the fact that $\inffpof{(\supp{m_1} \cup \supp{m_2})} \supseteq \supp{m_{12}}$. If support elements are not known, one can use the fact that $\inffpof{(\supp{m_1} \cup \supp{m_2})} = \inffpof{(\inffp{m_1} \cup \inffp{m_2})}$.
\end{example}

\subsection{\textbf{Generalized decompositions of evidence}}\label{dst_generalized_weights}

Some useful fusion rules for belief sources apply only to the conjunctive decomposition of evidence \cite{disj_dec_cautious_bold}. Such is the case for the Cautious conjunctive rule \cite{disj_dec_cautious_bold}, which is used when these sources are not independent. Yet, this decomposition can only be computed for mass functions $m$ such that $\Omega \in \supp{m}$. Here, we propose a generalization of this decomposition that works for any mass function $m$ such that $\bigcup \supp{m} \in \supp{m}$. This generalization is given by \\\autoref{generalized_dec}. A similar definition can be given for its dual, namely the disjunctive decomposition, for any mass function such that $\bigcap \supp{m} \in \supp{m}$.

\subsubsection{\textbf{Generalization}}

For any mass function $m$ such that $\Omega \in \supp{m}$, the conjunctive decomposition is defined as
\begin{align}\label{conj_dec}
m = \underset{{A\subset {\Omega}}}{\osymbol{\cap}} A^{w}
\end{align}
where each $A^{w}$ is a generalized simple mass function\footnote{We use abusively the term mass function in this article for the sake of simplicity. Actually, $A^{w}$ is a generalized simple basic belief assignment (GSBBA) (See \cite{conj_dec} for a more accurate terminology).}, defined as
\begin{align*}
\forall A \subset \Omega,~ \forall B \subseteq \Omega,\quad A^w(B) =
\begin{cases}
1 - w(A)	&\text{if $B = A$}\\
w(A)		&\text{if $B = \Omega$}\\
0 			&\text{otherwise}
\end{cases}
\end{align*}
and $w$ is the conjunctive weight function, i.e. the inverse of the multiplicative M\"obius transform of $q$ in $(2^\Omega, \supseteq)$, i.e. $w(A) = \displaystyle\prod_{\substack{B \subseteq \Omega\\B \supseteq A}} q(B)^{(-1)^{|B|-|A| +1}}$, where $q$ is the commonality function associated to $m$, i.e. the zeta transform of $m$ in $(2^\Omega, \supseteq)$.

In parallel, the M\"obius inversion theorem linking $q$ and $w$ gives us
\begin{align}\label{zeta_equality}
\forall y \in 2^\Omega,\quad q(y) = \sum_{x \supseteq y} m(y) = \prod_{x \supseteq y} w(y)^{-1},
\end{align}
which implies that $w(\Omega) = q(\Omega)^{-1} = m(\Omega)^{-1}$, hence $m(\Omega) \neq 0$. Even if we take the multiplicative M\"obius transform of $q$ instead of its inverse, we must have $m(\Omega) \neq 0$ so that the M\"obius inversion theorem can apply, as explained in section \ref{link_section}. Thus, in order to apply combination rules that are only defined on the conjunctive decomposition, such as the Cautious conjunctive rule, it was argued \cite{conj_dec} that one should only use mass functions $m$ satisfying $m(\Omega) \neq 0$. In practice, this is often done artificially, by discounting, i.e. multiplying all masses by some factor $\alpha \in (0,1)$ and assigning the complement to 1 to $m(\Omega)$. 

Nevertheless, assigning a mass to $\Omega$ is not ideal: when fusing by conjunction two mass functions that have different focal elements but $\Omega$ among them, it appears that no hypothesis (i.e. no focal element) is discarded in the resulting mass function. It contains the focal elements of the two original mass functions, as well as their pair-wise intersections. This means that hypotheses can only accumulate, making the number of focal elements (and so the number of focal points) explode in vast domains when many fusions of belief sources occur. Therefore, it is preferable to only assign masses to actual tangible hypotheses, instead of always considering that all hypotheses are possible even if unlikely. It is more stable, more accurate and allows for more use cases.

It turns out that it is possible to avoid the constraint $m(\Omega) \neq 0$ simply by using as weight function $w$ the inverse of the multiplicative M\"obius transform of $q$ in $({\downarrow \supp{m}}, \supseteq)$ instead of $(2^\Omega, \supseteq)$ (See section \ref{link_section}). But, how to reflect this in the conjunctive decomposition? Notice that Eq. \ref{conj_dec} is in fact equivalent to:
\begin{align}\label{conj_dec_dec}
\forall y \subseteq C,\quad q(y) =
\begin{cases}
\displaystyle\prod_{\substack{x \neq C}} w(x)	&\text{if $y = C$}\\
q(C) . \displaystyle\prod_{\substack{x \neq C\\x \supseteq y}} w(x)^{-1} &\text{otherwise}
\end{cases},
\end{align}
where $C=\Omega$, which means that all images of $q$ are determined by Eq. \ref{zeta_equality}, except for the one on $C$ which exploits the fact that the product of all weights is normalized to 1, due to Eq. \ref{zeta_equality} and Eq. \ref{constraint_m_alone}. Now, $C$ can be something else, as long as the conjunctive combination of simple mass functions as in Eq. \ref{conj_dec} leads to this form, which implies that $C$ both is less than or equal to the element replacing $\Omega$ in each simple mass function,  while not being less than $A$ (to satisfy the first line in Eq. \ref{conj_dec_dec}), and is a maximal element of the domain on which $w$ is defined (to satisfy $q(C) = w(C)^{-1}$, imposed by Eq. \ref{zeta_equality}). In addition, since the second line of this equation is only valid when $y$ is less than $C$, we get that $C$ must be above all elements of $\supp{m}$ for the conjunctive decomposition to account for all non-zero $q$ images defining $m$. Combining this with the fact that $w$ must be computed in $({\downarrow \supp{m}}, \supseteq)$, we get that $C = \bigcup\supp{m}$ and $C \in \supp{m}$, which is in accordance with the other aforementioned conditions.
This leads us to \autoref{generalized_dec}.
\begin{definition}\label{generalized_dec}
	For any mass function $m$ such that $\bigcup \supp{m} \in \supp{m}$, we define our generalized conjunctive decomposition as
	\begin{align*}
	m(B) = \begin{cases}
		\left(\underset{{A\subset {C}}}{\osymbol{\cap}} A_C^{w}\right)(B) &\text{if $B \subseteq C$}\\
		0 	&\text{otherwise}
	\end{cases}
	\end{align*}
	where $C = \bigcup \supp{m}$ and 
	\begin{align*}
	\forall A \subset C,~ \forall B \subseteq C,\quad A_C^w(B) =
	\begin{cases}
	1 - w^C(A)	&\text{if $B = A$}\\
	w^C(A)		&\text{if $B = C$}\\
	0 			&\text{otherwise}
	\end{cases}
	\end{align*}
	and $w^C(A) = \displaystyle\prod_{\substack{B \subseteq C\\B \supseteq A}} q(B)^{(-1)^{|B|-|A| +1}}$.
\end{definition}

If $\Omega \in \supp{m}$, then \autoref{generalized_dec} yields the classic conjunctive decomposition. This generalized decomposition is unique, as the classic one, due to the fact that $w^C$ is tied to $q$ by the M\"obius inversion theorem.

\subsubsection{\textbf{When $\supp{m}$ has no maximum}}\label{generalized_disc}

Similar to what was proposed in the past, it is possible to discount $m$ and assign the complement to 1 to any element between $\bigcup \supp{m}$ and $\Omega$, if $\bigcup \supp{m}$ is not already in $\supp{m}$.
We argue that $\bigcup \supp{m}$ should be chosen since it is the superset that supports the least hypotheses outside $\supp{m}$ and so the one that biases $m$ the least.

Furthermore, thanks to \autoref{emitfp} and \autoref{emi_link_fp_f_full}, we can even state the following \autoref{dec_discounting} about the resulting decomposition weights of such a discounting procedure.

\begin{corollary}\label{dec_discounting}
	For any mass function $m$ such that $\bigcup \supp{m} \not\in\supp{m}$, discounting it and assigning the complement to $\Omega$ gives the same decomposition weights as discounting it and assigning the complement to $\bigcup \supp{m}$.
\end{corollary}
\begin{proof}
	See Appendix \ref{appendix:dec_discounting}.
\end{proof}

\subsubsection{\textbf{Fusion of generalized decompositions}}\label{gen_dec_fusion}

Given \autoref{dec_discounting}, for two mass functions $m_1$ and $m_2$ such that neither $\supp{m_1}$ nor $\supp{m_2}$ has a maximum and $\bigcup \supp{m_1} = \bigcup \supp{m_2}$, discounting them and assigning the complement to $\Omega$ gives the same decomposition weights as discounting them and assigning the complement to $\bigcup \supp{m_1}$. Consequently, any fusion operator defined on decomposition weights produces the same results in our generalized decomposition as in the classic decomposition, except that the mass normally added to $\Omega$ is added to $\bigcup \supp{m_1}$ instead, which does not give credit to more hypotheses than needed and is more stable. In addition, this allows for new interesting cases when $\bigcup \supp{m_1} \neq \bigcup \supp{m_2}$. 

In this case, before applying any fusion operator, it is necessary to define a common domain for the resulting conjunctive decomposition. For combination rules based on the conjunction of two belief sources (i.e. both sources are considered reliable), such as Dempster's rule, the Cautious conjunctive rule, etc, domains must be intersected and evidence projected on this intersection. This common domain is simply $2^C$, where $C=\bigcup \supp{m_1} \cap \bigcup \supp{m_2}$. 
Then, concerning the projection of evidence on $2^C$, this consists in adding masses defined outside $2^C$ to their intersection with this domain. Fortunately, this is already what the commonality function $q$ does. Indeed, the projection of any mass function $m$ onto $2^C$ is obtained by transferring, i.e. adding, the mass on all elements $B \subseteq \Omega$ to the mass of $B \cap C$. This projection, noted $m_{\downarrow C}$, is itself a mass function since, by construction, it is as nonnegative as $m$ and the sum of its images remains unchanged. Notice that the zeta transform of $m_{\downarrow C}$ in $(2^C, \supseteq)$ is equal to $q$ on $2^C$ since $B \supseteq A$ is equivalent to $B \cap A = A$ and since for any element $A \subseteq C$, we have $A \cap (B \cap C) = A \cap B$. Thus, we can use the commonality functions $q_1$ and $q_2$, which are respectively the zeta transforms of $m_1$ in $(2^{T_1}, \supseteq)$ and $m_2$ in $(2^{T_2}, \supseteq)$, where $T_1 \supseteq {\bigcup \supp{m_1}}$ and $T_2 \supseteq {\bigcup \supp{m_2}}$. Then, it only remains to check that $q_1(C) \neq 0$ and $q_2(C) \neq 0$, which is equivalent to verifying that $C \in {\downarrow \supp{m_1}}$ and $C \in {\downarrow \supp{m_2}}$. The aforementioned discounting must be employed\footnote{Small but necessary approximation in order to apply the fusion operator.} on any commonality function that does not verify this condition. 
Now, we can finally fuse $m_1$ and $m_2$ by applying any conjunctive fusion operator to the weight functions $w^C_1$ and $w^C_2$ associated to respectively $q_1$ and $q_2$ in $(2^C, \supseteq)$. If \autoref{emitfp} is used to compute $w^C_1$ and $w^C_2$, notice that the focal points of $w^C_1$ are the pair-wise intersections $C \cap s_1$, where $s_1 \in\inffp{m_1}$, and the ones of $w^C_2$ are the pair-wise intersections $C \cap s_2$, where $s_2 \in \inffp{m_2}$.

The weight functions $w^C_1$ and $w^C_2$ correspond to ${m_1}_{\downarrow C}$ and ${m_2}_{\downarrow C}$, which are mass functions. Therefore, any fusion operator defined in the classic conjunctive decomposition, such as the Cautious conjunctive rule, is valid in this generalization, as if we had $\Omega = C$, and so can be applied to the weights of $w^C_1$ and $w^C_2$ (except the ones on $C$, in the same way that it is not applied to the weights on $\Omega$ in the classic decomposition). 

\begin{example}\label{ex:gen_dec2}
	Consider $\Omega = \{ a, b, c, d \}$ and let $m_1$ be a mass function such that $m_1(\{ a, b \}) = 0.2$, $m_1(\{ b, c \}) = 0.2$ and $m_1(\{ a \}) = 0.6$. We have $\bigcup\supp{m_1} = \{ a, b, c \}$ and $\inffp{m_1} = \supp{m_1} \cup \{ \{b\}, \emptyset \}$.
	Also, let $m_2$ be a mass function such that $m_2(\{ b, c \}) = 0.3$, $m_2(\{ c, d \}) = 0.1$ and $m_2(\{ c \}) = 0.6$. We have $\bigcup\supp{m_2} = \{ b,c,d \}$ and $\inffp{m_2} = \supp{m_2}$.
	Suppose we want to fuse $m_1$ and $m_2$ with the Cautious conjunctive rule. We pose $C=\bigcup\supp{m_1} \cap \bigcup\supp{m_2} =\{b,c\}$. We already have $C \in {\downarrow\supp{m_1}}$ and $C \in {\downarrow\supp{m_2}}$, so no discounting needed. We even have $C \in {\supp{m_1}}$ and $C \in {\supp{m_2}}$, which means that the focal points of $w^C_1$ are $2^C \cap \inffp{m_1} = \{ \{ b, c \}, \{ b \}, \emptyset \}$ and that the ones of $w^C_2$ are $2^C \cap \inffp{m_2} = \{ \{ b, c \}, \{ c \} \}$.
	Moreover, we have $q_1(\{b,c \}) = 0.2$, $q_1(\{b \}) = 0.4$ and $q_1(\emptyset) = 1$, where $q_1$ is the commonality function associated to $m_1$, and $q_2(\{b,c \}) = 0.3$ and $q_2(\{c \}) = 1$, where $q_2$ is the commonality function associated to $m_2$.
	We get:
	
	\noindent$w_1^C(\{b,c \}) = q_1(\{b,c\})^{-1} = \boldsymbol{5}\boldsymbol{, }$
	$w_1^C(\{b \}) = \left[q_1(\{b\}) ~.~ w_1^C(\{b,c \}) \right]^{-1} = \boldsymbol{0.5}\boldsymbol{, }$
	$w_1^C(\emptyset ) = \left[q_1(\emptyset) ~.~ w_1^C(\{b,c \}) ~.~ w_1^C(\{b \}) \right]^{-1} = \boldsymbol{0.4}$
	and 
	$w_2^C(\{b,c \}) = q_2(\{b,c\})^{-1} = \boldsymbol{\frac{1}{0.3}}\boldsymbol{, }$
	$w_2^C(\{c \}) = \left[q_2(\{c\}) ~.~ w_2^C(\{b,c \}) \right]^{-1} = \boldsymbol{0.3}$.
	
	Then, using the minimum operator\footnote{In fact, as all couples of numbers involve 1 and another number less than 1, the conjunctive rule would have given the same result here.} $\wedge$ of the Cautious conjunctive rule, we obtain:
	{	\small\begin{itemize}
			\item $w_{12}^C(\{b \}) = w_{1}^C(\{b \}) \wedge w_{2}^C(\{b \}) = 0.5 \wedge 1 = \boldsymbol{0.5}\boldsymbol{ }$
			\item $w_{12}^C(\{c \}) = w_{1}^C(\{c \}) \wedge w_{2}^C(\{c \}) = 1 \wedge 0.3 = \boldsymbol{0.3}\boldsymbol{ }$
			\item $w_{12}^C(\emptyset ) = w_{1}^C(\emptyset) \wedge w_{2}^C(\emptyset) = 0.4 \wedge 1 = \boldsymbol{0.4}\boldsymbol{ }$
		\end{itemize}
		\normalsize which means that $w_{12}^C(\{b,c\}) = \left[w_{12}^C(\{b \}) . w_{12}^C(\{c \}) . w_{12}^C(\emptyset)\right]^{-1} = \frac{1}{0.06}$.} Then, let us compute the associated commonality function $q_{12}$:
	$q_{12}(\{b,c \}) = w_{12}^C(\{b,c \})^{-1} = \boldsymbol{0.06}\boldsymbol{, }$
	$q_{12}(\{b \}) = \left[w_{12}^C(\{b \}) . w_{12}^C(\{b,c \})\right]^{-1} = \boldsymbol{0.12}\boldsymbol{, }$
	$q_{12}(\{c \}) = \left[w_{12}^C(\{c \}) . w_{12}^C(\{b,c \})\right]^{-1} = \boldsymbol{0.2}$ and 
	$q_{12}(\emptyset ) = \left[w_{12}^C(\emptyset) . w_{12}^C(\{b \}) . w_{12}^C(\{c \}) . w_{12}^C(\{b,c \})\right]^{-1} = \boldsymbol{1}$.
	
	Finally, the associated mass function $m_{12}$ is:
	$m_{12}(\{b,c \}) = q_{12}(\{b,c \}) = \boldsymbol{0.06}\boldsymbol{, }$
	$m_{12}(\{b \}) = q_{12}(\{b \}) - m_{12}(\{b,c \}) = \boldsymbol{0.06}\boldsymbol{, }$
	$m_{12}(\{c \}) = q_{12}(\{c \}) - m_{12}(\{b,c \}) = \boldsymbol{0.14}$ and 
	$m_{12}(\emptyset ) = q_{12}(\emptyset) - m_{12}(\{b \}) - m_{12}(\{c \}) - m_{12}(\{b,c \}) = \boldsymbol{0.74}\boldsymbol{, }$
	where $m_{12}(x) = 0$ for all $x \not \in {2^{\{b,c \}}}$. Since these results are here the same as if we had employed the conjunctive fusion rule, it is easy to see that they are correct. Indeed, computing the conjunctive fusion of $m_1$ and $m_2$ directly in $2^\Omega$ produces the same resulting mass function $m_{12}$. 
\end{example}

\subsection{\textbf{Better understanding the conjunctive and disjunctive decompositions}}\label{dst_ablation}

A better understanding of the conjunctive or disjunctive decomposition of evidence can be exploited to propose e.g. new fusion rules using \autoref{generalized_dec} or new approximation methods based on $\supp{w-1}$ instead of $\supp{m}$, where $m$ is a mass function and $w$ is the weight function associated to it.
For this, it is interesting 
to study how each piece of evidence in the conjunctive or disjunctive decomposition impacts the mass function. Until now, the only exact way to do this computation in the general case was using the FMT, which is $O(N.2^N)$, and this provided no insight on the internal workings of these modifications. Here, given \autoref{ablation_study_generalized}, we propose new formulas (\autoref{ablation_study_proposition}) describing the propagation of these updates. We will focus on the conjunctive decomposition, but a similar method exists for its dual, namely the disjunctive decomposition. Once focal points have been computed for the original weight function $w$, the complexity of our method ranges from $O(1)$ to $O(N~.~|\inffp{w-1}|)$ for each image of focal point modified in $w$, depending on its relation with respect to other focal points.

\begin{proposition}\label{ablation_study_proposition}
	Let $w'$ be equal to $w$ everywhere on $2^\Omega\backslash\{\Omega \}$, except for the image of some $x \in \inffp{w-1}\backslash \{\Omega \}$. Also, let $m'$ and $q'$ be the functions corresponding to $w'$ so that they satisfy Eq. \ref{link} in place of respectively $m$ and $q$. Then, we have:
	\begin{align}\label{ablation_study_q}
	\forall y \in 2^\Omega,\quad q'(y) = 
	\begin{cases}
	q(y) &\text{if $x \supseteq y$}\\
	\frac{w'(x)}{w(x)}.q(y) &\text{otherwise}
	\end{cases}
	\end{align}
	and 
	\begin{empheq}{align}\label{ablation_study_m}
		m'(y) = 
		\begin{cases}
			0 &\text{if $y \not\in \inffpof{S}$}\\
			\frac{w'(x)}{w(x)}.m(y) + \left[1-\frac{w'(x)}{w(x)}\right].m_{\downarrow x}(y)	&\text{if $x \supseteq y$}\\
			\frac{w'(x)}{w(x)}.m(y)	&\text{otherwise}
		\end{cases}
	\end{empheq}
	where $\inffpof{S} \supseteq \supp{w-1}\cup \{\Omega \}$ and $m_{\downarrow x} : {2^x} \rightarrow [0,1]$ is the mass function resulting from the projection of $m$ onto ${\downarrow x}$, i.e. the M\"obius transform of $q$ in $(2^x, \supseteq)$.
\end{proposition} 	
\begin{proof}
	See Appendix \ref{appendix:ablation_study_proposition}.
\end{proof}

\begin{example}\label{ex_ablation_1}
	Let us take back \autoref{ex_emt_2}, i.e. $\Omega = \{ a, b, c \}$ and $m$ is a mass function such that $m(\Omega) = 0.1$, $m(\{ a, b \}) = 0.1$, $m(\{ b, c \}) = 0.2$ and $m(\{ a \}) = 0.6$. We have $\supp{m} = \{\Omega, \{ a, b \}, \{ b, c \}, \{ a \} \}$ and $\inffp{m} = \supp{m} \cup \{ \{b\}, \emptyset \}$.
	We also have its commonality function $q$ such that $q(\Omega) = 0.1$, $q(\{a,b \}) = 0.2$, $q(\{b,c \}) = 0.3$, $q(\{a \}) = 0.8$, $q(\{b \}) = 0.4$ and $q(\emptyset) = 1$. Finally, its conjunctive weight function $w$ is defined by $w(\Omega) = 10$, $w(\{a,b \}) = 0.5$, $w(\{b,c \}) = \frac{1}{3}$, $w(\{a \}) = 0.25$, $w(\{b \}) = 1.5$ and $w(\emptyset) = 1.6$.
	
	Now, suppose that we would like to see how changing $w(\{b\})$ from 1.5 to 1, i.e. removing $\{b\}$ from $\supp{w-1}$, affects $m$. Then, Eq. \ref{ablation_study_m} tells us that only $m(\{b\})$ and $m(\emptyset)$ are impacted by more than just a renormalization factor. Let us start by computing $m_{\downarrow \{b\}}$, where ${\downarrow \{b\}} = 2^{\{b\}}$:
		$m_{\downarrow \{b\}}(\{b\}) = q(\{b\}) = \boldsymbol{0.4}$ and
		$m_{\downarrow \{b\}}(\emptyset) = q(\emptyset) - m_{\downarrow \{b\}}(\{b\}) = \boldsymbol{0.6}$.
	Now, we can update $m$ to provide $m'$:
{	\small\begin{itemize}
		\item $m'(\{b\}) = \frac{w'(\{b\})}{w(\{b\})}.m(\{b\}) + \left[1-\frac{w'(\{b\})}{w(\{b\})}\right].m_{\downarrow \{b\}}(\{b\}) = 0 + \left[1-\frac{1}{1.5}\right]*0.4 = \boldsymbol{\frac{2}{15}}$ 
		\item $m'(\emptyset) = \frac{w'(\{b\})}{w(\{b\})}.m(\emptyset) + \left[1-\frac{w'(\{b\})}{w(\{b\})}\right].m_{\downarrow \{b\}}(\emptyset) = 0 + \left[1-\frac{1}{1.5}\right]*0.6 = \boldsymbol{\frac{3}{15}}$
	\end{itemize}}
	All other focal points $p$ simply have their image through $m'$ renormalized:
	$$m'(p) = \frac{w'(\{b\})}{w(\{b\})}.m(p) = \frac{1}{1.5}.m(p)$$
	which gives $m'(\Omega) = m'(\{ a, b \}) = \boldsymbol{\frac{1}{15}}$, $m'(\{ b, c \}) = \boldsymbol{\frac{2}{15}}$ and $m'(\{ a \}) = \boldsymbol{\frac{6}{15}}$. Note that the sum of the images of $m'$ is indeed equal to 1.
\end{example}

\begin{example}\label{ex_ablation_2}
	Still using $m$, $q$ and $w$ from \autoref{ex_emt_2}, suppose now that we would like to see how changing $w(\{a,b\})$ from 0.5 to 1, i.e. removing $\{a,b\}$ from $\supp{w-1}$, affects $m$. Then, Eq. \ref{ablation_study_m} tells us that only $m(\{a,b\})$, $m(\{a\})$, $m(\{b\})$ and $m(\emptyset)$ are impacted by more than just a renormalization factor. Let us start by computing $m_{\downarrow \{a,b\}}$, where ${\downarrow \{a,b\}} = 2^{\{a,b\}}$:
{	\small\begin{itemize}
		\item $m_{\downarrow \{a,b\}}(\{a,b\}) = q(\{a,b\}) = \boldsymbol{0.2}\boldsymbol{ }$
		\item $m_{\downarrow \{a,b\}}(\{a\}) = q(\{a\}) - m_{\downarrow \{a,b\}}(\{a,b\}) = \boldsymbol{0.6}\boldsymbol{ }$
		\item $m_{\downarrow \{a,b\}}(\{b\}) = q(\{b\}) - m_{\downarrow \{a,b\}}(\{a,b\}) = \boldsymbol{0.2}$  
		\item $m_{\downarrow \{a,b\}}(\emptyset) = q(\emptyset) - m_{\downarrow \{a,b\}}(\{a,b\}) - m_{\downarrow \{a,b\}}(\{a\}) - m_{\downarrow \{a,b\}}(\{b\}) = \boldsymbol{0}$
	\end{itemize}
\normalsize	Now, we can update $m$ to provide $m'$: }
{	\small\begin{itemize}
		\item $m'(\{a,b\}) = \frac{w'(\{a,b\})}{w(\{a,b\})}.m(\{a,b\}) + \left[1-\frac{w'(\{a,b\})}{w(\{a,b\})}\right].m_{\downarrow \{a,b\}}(\{a,b\}) = \frac{1}{0.5}*0.1 + \left[1-\frac{1}{0.5}\right]*0.2 = \boldsymbol{0}$ 
		\item $m'(\{a\}) = \frac{w'(\{a,b\})}{w(\{a,b\})}.m(\{a\}) + \left[1-\frac{w'(\{a,b\})}{w(\{a,b\})}\right].m_{\downarrow \{a,b\}}(\{a\}) = \frac{1}{0.5}*0.6 + \left[1-\frac{1}{0.5}\right]*0.6 = \boldsymbol{0.6}$
		\item $m'(\{b\}) = \frac{w'(\{a,b\})}{w(\{a,b\})}.m(\{b\}) + \left[1-\frac{w'(\{a,b\})}{w(\{a,b\})}\right].m_{\downarrow \{a,b\}}(\{b\}) = 0 + \left[1-\frac{1}{0.5}\right]*0.2 =\boldsymbol{-0.2}$ 
		\item $m'(\emptyset) = \frac{w'(\{a,b\})}{w(\{a,b\})}.m(\emptyset) + \left[1-\frac{w'(\{a,b\})}{w(\{a,b\})}\right].m_{\downarrow \{a,b\}}(\emptyset) = 0 + \left[1-\frac{1}{0.5}\right]*0 = \boldsymbol{0}$
	\end{itemize}}
	All other focal points $p$ simply have their image through $m'$ renormalized:
	$$m'(p) = \frac{w'(\{b\})}{w(\{b\})}.m(p) = \frac{1}{0.5}.m(p)$$
	which gives $m'(\Omega) = \boldsymbol{0.2}$ and $m'(\{ b, c \}) = \boldsymbol{0.4}$. Notice that the sum of the images of $m'$ is indeed equal to 1, but we have a negative mass $m'(\{b\})$. This is expected as it is known (see \cite{conj_dec}) that the conjunctive decomposition is composed of what are called Simple Support Functions (SSF) (corresponding to elements of $\supp{w-1}$ with an image in $(0,1)$) and Inverse Simple Support Functions (ISSF) (corresponding to elements of $\supp{w-1}$ with an image in $(1, +\infty)$). Fusing all SSFs and ISSFs of one decomposition with the conjunctive rule gives the original mass function. SSFs correspond to mass functions, while ISSFs correspond to the decombination of mass functions, i.e. they are equivalent to mass functions with some negative value (if mass functions were allowed to have negative values). We had on $\{b\}$ a balance between positive mass values corresponding to $w(\{a,b \})$ and $w(\{b,c \})$ and a negative one corresponding to $w(\{b \})$, which we shifted towards the negatives by removing the positive mass corresponding to $w(\{a,b \})$. So, it just appears that changing SSFs from a decomposition that contains ISSFs is not always permitted in DST.
\end{example}

\section{\textbf{Conclusions and Perspectives}}\label{conclusion}

In this paper, we proposed an exact simplification of the zeta and M\"obius transforms, for any function in any incidence algebra or on any partially ordered set. From this, we introduced the notion of \textit{focal point} and discovered interesting properties when applied to the zeta and M\"obius transforms. Then, we applied our theorems to DST in order to allow for both exactitude and computational efficiency in vast domain for most transformations between representations of belief and for their fusion. We also proposed a generalization of the conjunctive decomposition of evidence and provided formulas uncovering the influence of each decomposition weight on the corresponding mass function. These last two applications demonstrate the potential of our approach for the proof of new theoretical results and may themselves be exploited to propose new discounting methods and fusion operators in DST.

To go further in practice, we need to present algorithms, data structures and experimental setups comparing execution times and memory usage between the EMT, the FMT and naïve approches using focal points.
Hence, we plan to issue a practical follow-up article in the near future, alongside a complete open-source implementation for DST.

In a more general way, our theoretical results can be both useful as a way to significantly reduce the complexity of any algorithm involving the zeta and M\"obius transforms (e.g. with our EMT \cite{me}, which is defined in any distributive lattice) and as tools to better understand them.

\bibliography{./complexity_reduction}

\begin{appendices}
	\section{Proofs}
	
	\subsection{\textbf{\autoref{emit}}}\label{appendix:emit}
	\begin{proof}
		For any set $S\supseteq \supp{f}$, given the fact that $\suppartof{S}$ partitions $P$ according to $S$ so that all elements of each part have the same lower closure in $S$,
		we have that all elements $x$ of a part $X$ in $\suppartof{S}$ share the same image $g_S(x) = \displaystyle\sum_{\substack{s \in S\\s \leq x}} f(s) = \displaystyle\sum_{\substack{s \in \supp{f}\\s \leq x}} f(s) + \displaystyle\sum_{\substack{s \in S\backslash\supp{f}\\s \leq x}} f(s) = g(x) + 0 = 
		g(X)$. Therefore, 
		it is possible to group the terms of the M\"obius inversion formula (Eq. \ref{mob_trans}) by parts of the level partition of any set $S \supseteq \supp{f}$, i.e. for any $y \in P$:
		\begin{align*}
		f(y) &= \sum_{\substack{x \leq y}}~ g(x)~.~\mu^{\phantom{\dagger}}_{P,\leq}(x,y)
		= \sum_{\substack{X \in \suppartof{S}\\{y \in \uparrow X}
		}} \sum_{\substack{z \in X\\z \leq y}} g(z)~.~ \mu^{\phantom{\dagger}}_{P,\leq}(z,y)\\
		&= \sum_{\substack{X \in \suppartof{S}\\{y \in \uparrow X}
		}} g(X) ~. \sum_{\substack{z \in X\\z \leq y}} \mu^{\phantom{\dagger}}_{P,\leq}(z,y)
		= \sum_{\substack{X \in \suppartof{S}\\{y \in \uparrow X}
		}}~ g(X)~.~ \eta^{\phantom{\dagger}}_{S, \leq, P}(X, y)
		\end{align*}
		Moreover, if there is a part $Z \in \suppartof{S}$ such that $\lowerclosurein{S}{Z} = \emptyset$, i.e. such that for all $z\in Z$ there is no element $s \in \supp{f}$ verifying $s \leq z$, then the image $g(Z)$ of this part is necessarily 0. Hence Eq. \ref{general_f}.
	\end{proof}

	\subsection{\textbf{\autoref{eta_min_theorem}}}\label{appendix:eta_min_theorem}

	\begin{proof}		
		For all $y \in P$, let us consider some $m \in \min\left(X\right)$ such that $m < y$. We can rewrite $\eta^{\phantom{\dagger}}_{S, \leq, P}(X, y)$ in the following form:
		\begin{align*}
		\eta^{\phantom{\dagger}}_{S, \leq, P}(X, y) &= \sum_{\substack{z \in X\\z \leq y}}\mu^{\phantom{\dagger}}_{P,\leq}(z,y)\nonumber
		= \sum_{\substack{z \in X\\z \leq y\\z\not\in {\uparrow m}}}\mu^{\phantom{\dagger}}_{P,\leq}(z,y) + \sum_{\substack{z \in X\\m \leq z \leq y}}\mu^{\phantom{\dagger}}_{P,\leq}(z,y)\\
		&= \sum_{\substack{z \in X\\z \leq y\\z\not\in {\uparrow m}}}\mu^{\phantom{\dagger}}_{P,\leq}(z,y) + \hspace{-0.15cm}\sum_{\substack{m \leq z \leq y}}\mu^{\phantom{\dagger}}_{P,\leq}(z,y) - \hspace{-0.15cm}\sum_{\substack{z \not\in X\\m \leq z \leq y}}\mu^{\phantom{\dagger}}_{P,\leq}(z,y)
		\end{align*}
		which reduces to:
		\begin{align}\label{eta_split}
		\eta^{\phantom{\dagger}}_{S, \leq, P}(X, y)
		&= \epsilon^{\phantom{\dagger}}_{m}(X,y) - \sum_{\substack{z \not\in X\\m \leq z \leq y}}\mu^{\phantom{\dagger}}_{P,\leq}(z,y)
		\end{align}
		where $\epsilon^{\phantom{\dagger}}_{m}(X,y) = \hspace{-0.15cm}\displaystyle\sum_{\substack{z \in X\\z \leq y\\z\not\in {\uparrow m}}}\mu^{\phantom{\dagger}}_{P,\leq}(z,y)$, since Eq. \ref{mob_func_sum} gives us $\hspace{-0.15cm}\displaystyle\sum_{\substack{m \leq z \leq y}}\mu^{\phantom{\dagger}}_{P,\leq}(z,y) = 0$.
		Now, let us determine from which parts the elements of $\{z \not\in X ~/~ m \leq z \leq y\}$ are. First, any of these elements belongs to a part $Z$ from $\suppartof{S}\backslash \{X\}$. Then, any element $z$ of these $Z$ parts satisfies $m \leq z$. 
		So, 
		we have ${\downarrow m} \subseteq {\downarrow z}$,
		which implies that $\lowerclosurein{S}{X} \subseteq \lowerclosurein{S}{Z}$ by definition of a level partition. More precisely, we get $\lowerclosurein{S}{X} \subset \lowerclosurein{S}{Z}$, since $Z \neq X$.
		Thus, the only parts $Z \in \suppartof{S}$ that need to be considered in the sum of Eq. \ref{eta_split} satisfy $\lowerclosurein{S}{X} \subset \lowerclosurein{S}{Z}$. 
		
		From this, without any hypothesis on the level partition, we can show that $\eta$ can be written in a recursive form, but not without $\epsilon$, which can only be computed from $\mu$.
		However, if we consider that every part $Z \in \suppartof{S}$ satisfying $\lowerclosurein{S}{X} \subseteq \lowerclosurein{S}{Z}$ has a minimum, then we can get rid of $\epsilon$. 
		Indeed, if $X$ has a minimum, then every element of $X$ is greater than $m$, i.e. $\epsilon^{\phantom{\dagger}}_{m,\leq}(X,y) = \sum_{\substack{z \in X\\z \leq y\\z\not\in {\uparrow m}}}\mu^{\phantom{\dagger}}_{P,\leq}(z,y) = 0$. The same goes for the rest of the recursion, hence the need for the parts $\lowerclosurein{S}{X} \subset \lowerclosurein{S}{Z}$ to each have a minimum too. 
		
		Moreover, these conditions mean that their respective lower closures in $S$ all have a supremum, since the supremum is by definition the least upper bound. Thus, we get that the only parts $Z \in \suppartof{S}$ that need to be considered in the sum of Eq. \ref{eta_split} satisfy $\bigvee \lowerclosurein{S}{X} < \bigvee \lowerclosurein{S}{Z}$, where $\bigvee \lowerclosurein{S}{X} = m = \bigwedge X$ and $\bigvee \lowerclosurein{S}{Z} = \bigwedge Z$. So, we have:
		\begin{align*}
		\eta^{\phantom{\dagger}}_{S, \leq, P}(X, y)
		&= 
		- \sum_{\substack{z \not\in X\\m \leq z \leq y}}\mu^{\phantom{\dagger}}_{P,\leq}(z,y)
		=  
		- \sum_{\substack{Z \in \suppartof{S}\\m < \bigwedge Z \leq y}} \sum_{\substack{z \in Z\\m \leq z \leq y}}\mu^{\phantom{\dagger}}_{P,\leq}(z,y)\\
		&= - \sum_{\substack{Z \in \suppartof{S}\\\bigwedge X < \bigwedge Z \leq y}} \sum_{\substack{z \in Z\\z \leq y}}\mu^{\phantom{\dagger}}_{P,\leq}(z,y)
		= - \sum_{\substack{Z \in \suppartof{S}\\\bigwedge X < \bigwedge Z \leq y}} \eta^{\phantom{\dagger}}_{S, \leq, P}(Z, y)
		\end{align*}
		In addition, we know from \autoref{eta_def} that if $y \in \min\left(X\right)$, i.e. $y = \bigwedge X$ if $X$ has a minimum, then we have
		$\eta^{\phantom{\dagger}}_{S, \leq, P}(X, y) = 1$.
	\end{proof}

	\subsection{\textbf{\autoref{fp_from_g}}}\label{appendix:fp_from_g}
	
	\begin{proof}
		{
			For the sake of clarity, we will use the alias $S = \supp{f}$ in the following.
			Let $G$ be the set made of the minimal elements of each part of the partition $\mathcal{G}$ defining $g$, i.e. $G = \bigcup_{X \in \mathcal{G}} \min(X)$. We will assume that $\supfpof{G}$ is an upper subsemilattice of $P$. Let us now find the elements of $\supp{f}$ with $G$.}
		
		{(i) It is obvious that $\min(P) \subseteq G$. 
		(ii) For any elements $y\in P$ and $s \in \supp{f}$ such that $y < s$ and $|\lowerclosurein{S}{s}\backslash\lowerclosurein{S}{y}| = 1$, we have $g(s) = g(x) + f(s) \neq g(x)$, for any element $x \in P$ where $y \leq x < s$. Therefore, there exists a part $X \in \mathcal{G}$ such that $s \in \min(X)$, i.e. $s \in G$.}
		{
			(iii) For any elements $y \in G$ and $s \in \supp{f}$ such that $y < s$ and $|\lowerclosurein{S}{s}\backslash\lowerclosurein{S}{y}| = n$, where $n \geq 2$, there exists an element $s' \in \supp{f}$ such that $s' < s$ and $s' \not\leq y$. 
		Thus, there is an element $y' = y \vee s'$ where $y' \leq s$ and $|\lowerclosurein{S}{s}\backslash\lowerclosurein{S}{y'}| \leq n-1$. 
		If $y' \neq s$ and $|\lowerclosurein{S}{s}\backslash\lowerclosurein{S}{y'}| \geq 2$, then there is an element $s'' \in \supp{f}$ such that $s'' < s$ and $s'' \not\leq y'$, which means that there is an element $y'' = y' \vee s''$ verifying both $|\lowerclosurein{S}{s}\backslash\lowerclosurein{S}{y''}| \leq n-2$ and $y'' \leq s$. This upward recursion will ultimately end either because the number of elements in the difference of lower closures in $S$ reaches 1 (which means that 
		$s \in G$, given (ii)) or because it has generated $s$ with the supremum of two lower elements.
			
		Furthermore, from what we demonstrated up to this point, either the aforementionned element $s'\in \supp{f}$ 
		is in $G$ (by (i) or (ii)) or there exists an element $x \in G$ such that $x < s'$ verifying $|\lowerclosurein{S}{s'}\backslash\lowerclosurein{S}{x}| \geq 2$,
	 which means that there exists an element $\tilde{s} \in \supp{f}$ such that 
	 $\tilde{s} < s'$ and $\tilde{s} \not\leq x$.
	 And again, either $\tilde{s} \in G$ or there exists an element $x' \in G$ such that $x' < \tilde{s}$ verifying $|\lowerclosurein{S}{\tilde{s}}\backslash\lowerclosurein{S}{x'}| \geq 2$.
	 Noticing that $|\lowerclosurein{S}{\tilde{s}}| < |\lowerclosurein{S}{s'}| < |\lowerclosurein{S}{s}|$, we know that this downward recursion will ultimately end with an element $\dot{s} \in \supp{f}$ that is in $G$ since the lower closure in $S$ decreases and will eventually contain only one element of $S$. Therefore, either $s'$ is in $G$ or it can be found by computing successive suprema as in (iii), starting with the supremum between $\dot{s}$ and another element of $G$. The same goes for $s''$ and any other elements eventually encountered in the upward recursion of (iii). Hence, all these elements can be used with $y$ to find $s$ again through successive suprema.
	 So, we get that all elements of $\supp{f}$ are in the join-closure of $G$, i.e. $\supfpof{G}$.} 
		
		We can be even more precise than this by noticing that the supremum of two minimal elements of $P$ associated with 0 through $g$ cannot be an element of $\supp{f}$ that is not already in $G$. Indeed, for any element $m \in \min(P)$, we have $g(m)=f(m)$. So, if $g(m)=0$, then $f(m)=0$, i.e. $m \not\in \supp{f}$. Yet, each supremum in (iii) involves at least one element of $\supp{f}$, i.e. one element that is not both a minimal element and an element with null image through $g$. This means that the supremum of two elements of $M$, where $M = \min(P) \cap \overline{\supp{g}}$, if not already in $G$, is either not in $\supp{f}$ or the supremum of another set of elements, where at least one is in $G\backslash M$.
		Thus, we would like to avoid computing any supremum of the form $\bigvee A$, where $A \subseteq M$, 
		in our join-closure. 
		Let $Y$ be the set made of every supremum $x \vee a$, where $x \in G\backslash M$ and $a \in M$. 
		Notice that the elements of $\supfpof{(Y \cup G\backslash M)}$ are of the form $\bigvee \left(X \cup A \right)$, where $\emptyset \subset X \subseteq G\backslash M$ and $A \subseteq M$.
		Therefore, $\supfpof{(Y \cup G\backslash M)}$ contains all the elements of $\supfpof{G}$, except the ones that are exclusively of the form $\bigvee A$, where $A \subseteq M$.
		Consequently, we have $\supp{f} \subseteq \supfpof{(Y \cup G\backslash M)}$, which means, by definition of a closure operator, that $\supfpof{\supp{f}} \subseteq \supfpof{(Y \cup G\backslash M)} \subseteq \supfpof{G}$.
	\end{proof}

	\subsection{\textbf{\autoref{emi_link_fp_f_full}}}\label{appendix:emi_link_fp_f_full}
	
	\begin{proof}
		Let $\mathcal{G}^*$ be the partition of $P$ w.r.t. the images of $g$ such that $\mathcal{G}^* = \suppartof{\supp{f}}$, and let $G^* = \bigcup_{X \in \mathcal{G}^*} \min(X)$.
		We have $G^* = \supfpof{\supp{f}} \cup \min(P)$, since $\min(P)$ contains the minimal elements of the only part of $\suppartof{\supp{f}}$ that is not in ${\uparrow\supp{f}}$, if it even exists. 
		Adapting \autoref{fp_from_g} to the multiplicative M\"obius transform, we also have $\supfpof{\supp{h-1}} \subseteq \supfpof{G^*} = \supfpof{(\supp{f} \cup \min(P))}$, which means that $\supfpof{(\supp{h-1} \cup \min(P))} \subseteq \supfpof{(\supp{f} \cup \min(P))}$.
		The same reasoning can be applied to the partition $\mathcal{G}_h^* = \suppartof{\supp{h-1}}$, leading to $\supfpof{(\supp{f} \cup \min(P))} \subseteq \supfpof{(\supp{h-1} \cup \min(P))}$.
		Therefore, combining all inequalities, we get that $\supfpof{(\supp{f} \cup \min(P))} = \supfpof{(\supp{h-1} \cup \min(P))}$. Finally, by \autoref{supp_zero}, we have $\supfpof{\supp{f}} = \supfpof{(\supp{h-1} \cup \min(P))}$.
	\end{proof}

\subsection{\textbf{\autoref{ablation_study_generalized}}}\label{appendix:ablation_study_generalized}

\begin{proof}
	In formal terms, the condition on $h'$ translates for any $y\in P$ to:
	\begin{align*}
	h'(y) = \begin{cases}
	h'(y)	&\text{if $y = x$}\\
	h(y)	&\text{otherwise}
	\end{cases}
	\end{align*}
	By Eq. \ref{link}, we have that for any $y \in P$,
	$
	g'(y) = \displaystyle\prod_{\substack{z \leq y}} h'(z).
	$
	We observe that $h'(x)$ appears in the product of $g'(y)$ only if $x\leq y$. Since all other images of $h'$ are equal to the ones of $h$, we get Eq. \ref{influence_w_g}. 
	{Then, \autoref{emi_link_fp_f_full} gives us $\supfpof{\supp{f}} = \supfpof{(\supp{h-1} \cup \min(P))}$ and $\supfpof{\supp{f'}} = \supfpof{(\supp{h'-1} \cup \min(P))}$. Moreover, notice that $\supp{h'-1} \subseteq \supp{h-1} \cup \{ x \}$. We get $\supfpof{(\supp{h'-1} \cup \min(P))} \subseteq \supfpof{(\supp{h-1} \cup \min(P) \cup \{ x \})}$, which implies that $\supfpof{\supp{f'}} \subseteq \supfpof{(\supp{f} \cup \{ x \})}$.}
	Consequently, we know that any upper subsemilattice $\supfpof{S}$ of $P$ such that $\supfpof{S} \supseteq \supp{f} \cup \{ x \}$ can be used in Eq. \ref{general_f_supfp} of \autoref{emitfp} for both $f$ and $f'$. So, we get 
	for any $y \in P$:
	\begin{align*}
	f'(y) 
	&= \sum_{\substack{s \in \supfpof{S}\\s \leq y}} \hspace{-0.cm}g'(s)~.~ \eta^{\phantom{\dagger}}_{S, \leq, P}(s, y)\nonumber\\
	&= \sum_{\substack{s \in \supfpof{S}\\x \leq s \leq y}} \hspace{-0.cm}g(s)~.~\frac{h'(x)}{h(x)}~.~ \eta^{\phantom{\dagger}}_{S, \leq, P}(s, y) + \sum_{\substack{s \in \supfpof{S}\\s \leq y\\s\not\in {\uparrow x}}} \hspace{-0.cm}g(s)~.~ \eta^{\phantom{\dagger}}_{S, \leq, P}(s, y)\nonumber
	\end{align*}
	\begin{align*}
	\phantom{f'(y)} &= \frac{h'(x)}{h(x)}~.~\sum_{\substack{s \in \supfpof{S}\\x \leq s \leq y}} \hspace{-0.cm}g(s)~.~ \eta^{\phantom{\dagger}}_{S, \leq, P}(s, y) + f(y) - \sum_{\substack{s \in \supfpof{S}\\x \leq s \leq y}} \hspace{-0.cm}g(s)~.~ \eta^{\phantom{\dagger}}_{S, \leq, P}(s, y)\nonumber\\
	&= f(y) + \left[\frac{h'(x)}{h(x)}-1\right].\sum_{\substack{s \in \supfpof{S}\\x\leq s \leq y}} g(s)~.~\eta^{\phantom{\dagger}}_{S, \leq, P}(s, y)
	\end{align*}
	
	Besides, 
	for any $s,y \in P$ where $x \leq s < y$, we have $\{p ~/~ s < p \} = \{p ~/~ x \leq s < p \}$ and so:
	\begin{align*}
	\eta^{\phantom{\dagger}}_{S, \leq, P}(s, y) &= - \sum_{\substack{p \in \supfpof{S}\\s < p \leq y}} \eta^{\phantom{\dagger}}_{S, \leq, P}(p, y)
	= - \sum_{\substack{p\in {\uparrow x}\\p \in \supfpof{S}\\s < p \leq y}} \eta^{\phantom{\dagger}}_{S, \leq, P}(p, y)
	= \eta^{\phantom{\dagger}}_{S, \leq, \uparrow x}(s, y),
	\end{align*}
	where ${\uparrow x}$ is the upper closure of $x$ in $P$.
	Therefore, the expression $\displaystyle\sum_{\substack{s \in \supfpof{S}\\x\leq s \leq y}} g(s)~.~\eta^{\phantom{\dagger}}_{S, \leq, P}(s, y)$ is in fact the M\"obius transform of $g$ in $({\uparrow x}, \leq)$.
\end{proof}

\subsection{\textbf{\autoref{dec_discounting}}}\label{appendix:dec_discounting}

\begin{proof}
	For any mass function $m$ such that $C \in \supp{m}$, where $C = \bigcup \supp{m}$, \autoref{emi_link_fp_f_full} implies that $\inffp{m}\backslash\{C \} = \inffp{w^C-1}\backslash\{C \}$, where $w^C$ is the inverse of the multiplicative M\"obius transform of $q$ in $({\downarrow C}, \supseteq)$ and $q$ is the zeta transform of $m$ in $({\downarrow T}, \supseteq)$, where $T \supseteq C$. In particular, suppose $C\neq \Omega$ and let $w'$ be the inverse of the multiplicative M\"obius transform of $q'$ in $(2^\Omega, \supseteq)$, where $q'$ corresponds to $m'$ and $m'$ is equal to $m$ everywhere, except that $m'(C) = 0$ and $m'(\Omega) = m(C)$. We have $\inffp{w'-1} \backslash\{\Omega \} = \inffp{m'}\backslash\{\Omega \} = \inffp{m}\backslash\{C \} = \inffp{w^C-1}\backslash\{C \}$.
	
	Furthermore, $q'(\Omega) = q(C)$ and for any element $y \subseteq C$, we have $q'(y) = q(y)$, which implies that for any element $s \in \inffpof{\supp{w^C-1}}\backslash\{C\}$, we have $q'(y) = q(y)$. This means that $w^C(C) = q(C)^{-1} = w'(\Omega)$ and that by \autoref{emitfp}, for any element $y \subset C$, we have $w^C(y) = \displaystyle\prod_{\substack{s \in \inffpof{S}\\s \supseteq y}} q(s)^{-\eta^{\phantom{\dagger}}_{S, \supseteq, {\downarrow a}}(s, y)} = \displaystyle\prod_{\substack{s \in \inffpof{S'}\\s \supseteq y}} q'(s)^{-\eta^{\phantom{\dagger}}_{S', \supseteq, P}(s, y)} = w'(y)$, where $S= \supp{w^C-1}$ and $S'= S\backslash\{C\} \cup \{\Omega \}$ and $\eta^{\phantom{\dagger}}_{S, \supseteq, {\downarrow C}}(C, y) = \eta^{\phantom{\dagger}}_{S', \supseteq, P}(\Omega, y)$, since $C$ and $\Omega$ have the same relations with respect to the elements in $\inffp{w^C-1}\backslash\{C\}$ and so the same role in the recursion giving $\eta$ in Eq. \ref{eta_supfp}. 
\end{proof}

\subsection{\textbf{\autoref{ablation_study_proposition}}}\label{appendix:ablation_study_proposition}

	\begin{proof}
		Let $w'$ be equal to $w$ everywhere on $2^\Omega$, except for the image of some $x \in \inffp{w-1}\backslash \{\Omega \}$. Also, let $m'$ and $q'$ be the functions corresponding to $w'$ so that they satisfy Eq. \ref{link} in place of respectively $m$ and $q$. For any element $y \in 2^\Omega$,
		\autoref{ablation_study_generalized} gives us:
		\begin{align*}
		q'(y) = 
		\begin{cases}
		\frac{w(x)}{w'(x)}.q(y) &\text{if $x \supseteq y$}\\
		q(y) &\text{otherwise}
		\end{cases}
		\end{align*}
		and 
		\begin{align}\label{ablation_study_m_unnormalized}
		m'(y) = 
		\begin{cases}
		0 &\text{if $y \not\in \inffpof{S}$}\\
		m(y) + \left[\frac{w(x)}{w'(x)}-1\right].m_{\downarrow x}(y)	&\text{if $x \supseteq y$}\\
		m(y)	&\text{otherwise}
		\end{cases},
		\end{align}
		where $\inffpof{S} \supseteq \supp{m}$ and $m_{\downarrow x} : {\downarrow x} \rightarrow \mathbb{R}$ is the M\"obius transform of $q$ in $({\downarrow x}, \supseteq)$.
		However, in DST, we have to respect a normalization constraint on $m'$ and $w'$ that is imposed by Eq. \ref{constraint_m_alone} and Eq. \ref{zeta_equality}.
		Obviously, looking at the normalized product, we see that the only way to normalize $w'$ without changing any other image of $2^\Omega\backslash\{\Omega \}$ through $w$ is by changing the one on $\Omega$. More precisely, the normalized equivalent $\overline{w'}$ of $w'$ must satisfy $\overline{w'}(\Omega) = \frac{w(x)}{w'(x)} . w(\Omega)$. 
		So, taking back Eq. \ref{ablation_study_m_unnormalized}, but this time updating $m'$ to get its normalized equivalent $\overline{m'}$ with $x=\Omega$, which implies that $\downarrow x = {\downarrow \Omega} = 2^\Omega$, we have for any $y \in 2^\Omega$:
		\begin{align*}
		\overline{m'}(y) &= m'(y) + \left[\frac{w'(\Omega)}{\overline{w'}(\Omega)}-1\right].m'(y)
		= \frac{w'(\Omega)}{\overline{w'}(\Omega)}.m'(y) 
		= \frac{w'(x)}{w(x)}~.~m'(y)
		\end{align*}
		Discarding $m'$ as it is not a mass function (does not satisfy Eq. \ref{constraint_m_alone}), we consider directly $\overline{m'}$ as the result of the modification of one image in the conjunctive decomposition. Hence Eq. \ref{ablation_study_q} and Eq. \ref{ablation_study_m} with
		$\inffpof{S} \supseteq \supp{m}$.
		In addition, since $2^\Omega$ has a maximum $\Omega$, we have $\inffp{m} = \inffp{w-1} \cup \{\Omega \}$ by \autoref{emi_link_fp_f_full}. 
		
		Finally, remark that $m_{\downarrow x}$ is a mass function, since it is the M\"obius transform, in a reduced subset, of $q$, which corresponds to the mass function $m$. It is the same projection process as the one producing $m_{\downarrow C}$ in section \ref{gen_dec_fusion}.
	\end{proof}
\end{appendices}

\end{document}